# Quasi-periodic oscillations, charge and the gravitomagnetic theory

Jacob Biemond*

*Vrije Universiteit, Amsterdam, Section: Nuclear magnetic resonance, 1971-1975*
[*]*Postal address: Sansovinostraat 28, 5624 JX Eindhoven, The Netherlands*
Website: http://www.gewis.nl/~pieterb/gravi/    Email: gravi@gewis.nl

## ABSTRACT

A new model for the explanation of the high frequency quasi-periodic oscillations (QPOs) in pulsars, black holes and white dwarfs is presented. Three circular tori are assumed to be present around the star: an inner torus with charge $Q_i$, an outer torus with charge $Q_o$ and a torus with electrically neutral mass $m_m$ in the middle, whereas the star bears a charge $Q_s$ ($Q_o$ and $Q_s$ have the same sign, $Q_i$ the opposite one). The frequency $v_m$ of the mass current is approximately given by the Kepler frequency, whereas the frequencies of $Q_i$ and $Q_o$, $v_i$ and $v_o$, respectively, are calculated from classical mechanics and Coulomb's law.

For the explanation of the low frequency QPOs in pulsars and black holes a special interpretation of the gravitomagnetic theory may be essential. From the latter theory four new gravitomagnetic precession frequencies are deduced, which may be identified with the observed low frequency QPOs.

Predictions of the presented model are compared with observed high frequency and low frequency QPOs of the pulsars SAX J1808.4–3658, XTE J1807–294, IGR J00291+5934, SGR 1806–20 and the black hole XTE J1550–564. In addition, charge flow near the pole of pulsars may explain frequency drift of burst oscillations. Moreover, charge flow at the equator of SAX J1808.4–3658 may be the cause of the enigmatic 410 kHz QPO. Furthermore, the Lense-Thirring frequency is discussed and a modified formula is compared with data of the pulsars.

Contrary to pulsars and black holes, the low frequency QPOs of white dwarfs might be attributed to electromagnetic precession frequencies, deduced in this work. Predictions are compared with data of the dwarf nova VW Hyi.

Summing up, the new model seems to be in agreement with more observations than previously proposed alternatives.

## 1. INTRODUCTION

Quasi-periodic oscillations (QPOs) have been observed for many accreting pulsars, black holes and white dwarfs. Data from pulsars and black holes have been reviewed by, e.g., van der Klis [1]. Usually, a distinction is made between high frequency quasi-periodic oscillations and a low frequency complex. A survey of different types of periodic oscillations for accreting white dwarfs has been given by Warner, Woudt and Pretorius [2]. They distinguished three types: short-period dwarf nova oscillations (DNOs), long-period DNOs (lpDNOs) and quasi-periodic oscillations (QPOs) with a very long period.

QPOs are important from the theoretical point of view, for they may originate near the surface of the star, where effects predicted by general relativity may become manifest. Numerous models have been proposed to explain the origin of QPOs (see for a review, e.g., [1, § 2.8]). For example, a relativistic precession model was proposed by Stella and Vietri [3] and Stella, Vietri and Morsink [4]. In the latter model three formulas for the QPO frequencies were considered. The QPO with the highest frequency, the so-called *upper* kHz QPO, $v_u$, was identified with an orbital frequency of neutral particles, closely related to the Kepler frequency, whereas the *lower* kHz QPO, $v_l$, was attributed to the so-called periastron precession frequency. A third lower frequency QPO was identified with the Lense-Thirring frequency. In spite of numerous attempts, however, a generally accepted model to explain all properties of the observed QPOs has not yet emerged.

Compared with other models, we introduce a new key ingredient: *charge*. The high frequency QPOs are attributed to orbital frequencies of circular tori around the central star. Three different high frequency QPOs will be distinguished: the first frequency $v_i$ is attributed to an *inner* torus containing a total electric charge $Q_i$. The sign of the charge of $Q_i$ is assumed to be opposite to the sign of the total charge $Q_s$ of the *star*. The second frequency $v_o$ is attributed to an *outer* torus with total charge $Q_o$ ($Q_o$ and $Q_s$ have the same sign). Further, a third torus, containing a total electrically neutral mass $m_m$ (the subscript m stems from *middle*) is assumed to be present between the two other tori. The latter torus generates a third frequency, $v_m$, closely related to the Kepler frequency.

By using classical mechanics and Coulomb's law, it will be shown in section 2 that equilibrium between the charge $Q_s$ of the star and the charges $Q_o$ and $Q_i$ in the tori is possible. It appears that equilibrium is only possible, if the angle between the planes of the tori is not too large. This equilibrium condition may (partially) explain the observed instability of the quasi-periodic oscillations.

For the interpretation of the low frequency QPOs of pulsars and black holes another key ingredient may be essential, i.e., *a special interpretation of the gravito-magnetic theory*, which may be deduced from general relativity [5–7]. In this version the so-called "magnetic-type" gravitational field is identified as a common magnetic field. The latter identification leads to a prediction of the magnetic field of the star. This prediction will be used in the derivations below. Therefore, we first pay some attention to this consequence of our interpretation.

Identification of "magnetic-type" gravitational field with a magnetic field results into the so-called Wilson-Blackett formula. This relation applies, for example, to a spherical star consisting of electrically neutral matter

$$\mathbf{M}(\text{gm}) = -\tfrac{1}{2}\beta c^{-1} G^{\frac{1}{2}} \mathbf{S}. \qquad (1.1)$$

Here $\mathbf{M}(\text{gm})$ is the *(gravito)*magnetic dipole moment of the star with angular momentum $\mathbf{S}$, $c$ is the velocity of light in vacuum, $G$ is the gravitational constant and $\beta$ is a dimensionless constant of order unity. Available observations and theoretical considerations with respect to the relation (1.1), and other explanations of the origin of the magnetic field of celestial bodies have been reviewed by Biemond [6]. The magnetic field of pulsars has separately been discussed [8]. The angular momentum $\mathbf{S}$ for a spherical star with a total mass $m_s$ and radius $r_s$ can be found from the relations

$$\mathbf{S} = I\mathbf{\Omega}_s, \text{ or } S = I\Omega_s = \tfrac{2}{5} f_s m_s r_s^2 \Omega_s, \qquad (1.2)$$

where $\Omega_s = 2\pi v_s$ is the angular velocity of the star ($v_s$ is its rotational frequency), $I$ is the moment of inertia of the star and $f_s$ is a dimensionless factor depending on the homogeneity of the mass density in the star (for a homogeneous mass density $f_s = 1$). For convenience sake, a value $f_s = 1$ will be assumed in this work.

The value of a gravitomagnetic dipole moment $\mathbf{M}$ (or an electromagnetic dipole moment $\mathbf{M}$) can be calculated from

$$\mathbf{M} = \tfrac{1}{2} R^3 \mathbf{B}_p, \text{ or } M = \tfrac{1}{2} R^3 B_p. \qquad (1.3)$$

Here $\mathbf{B}_p$ is the magnetic induction field at, say, the north pole of the star at distance $R$ from the centre of the star to the field point where $\mathbf{B}_p$ is measured ($R \geq r_s$).

Combination of (1.1)–(1.3) yields the following gravitomagnetic prediction for $\mathbf{B}_p$

$$\mathbf{B}_p(\text{gm}) = -\tfrac{2}{5}\beta c^{-1} G^{\frac{1}{2}} m_s r_s^{-1} \mathbf{\Omega}_s. \qquad (1.4)$$

The minus sign reflects that the vectors $\mathbf{B}_p(gm)$ and $\mathbf{\Omega}_s$ possess opposite directions for $\beta = +1$. Neither the sign nor the value of $\beta$ follows from the gravitomagnetic theory. It is stressed that $\mathbf{B}_p(gm)$ at distance $r_s$ has been derived for an ideal gravitomagnetic dipole located at the centre of the star. However, the same result for $\mathbf{B}_p(gm)$ can be deduced for a homogeneous mass distribution in the star [9].

If the star bears a total charge $Q_s$, an additional *(electro)*magnetic dipole moment $\mathbf{M}(Q_s)$ is also present

$$\mathbf{M}(Q_s) = \tfrac{1}{2} c^{-1} m^{-1} Q_s \mathbf{S}. \tag{1.5}$$

This relation is analogous to (1.1). High values for the charge $Q_s$ in stars have been considered by several authors, e.g., by Ghezzi [10]. He discussed values of $Q_s$ up to the extreme limit $|Q_s| = G^{\frac{1}{2}} m_s$, corresponding to a mixture of $1.11 \times 10^{18}$ neutrons and one proton, for example. The Coulomb repulsion is reduced for a homogeneous charge distribution. However, the stability of stars with such large amounts of charge is a matter of debate (see, e.g., refs. [10] and [8]).

In this work it is shown that an equilibrium is possible between a charge $Q_s$ in the star and the charges $Q_i$ and $Q_o$ in the tori. Especially, accretion will affect this equilibrium between the charges $Q_s$, $Q_i$ and $Q_o$. Furthermore, it is noticed that the adopted charge $Q_s$ may lead to strong electric fields of order of the critical electric field $E_c = m_e^2 c^3 / e\hbar = 4.414 \times 10^{13}$ statvolt.cm$^{-1}$ = $1.323 \times 10^{18}$ V. m$^{-1}$. At this field strength electron-positron pair creation may become probable. However, this subject will not be treated in this paper.

As already stated, for the interpretation of the low frequency QPOs, another consequence of the applied gravitomagnetic theory may be essential. The latter theory predicts an angular precession velocity $\mathbf{\Omega}(gm)$ of any angular momentum $\mathbf{S}$, for example, the angular momentum of a star or a torus. The following relation then applies to $\mathbf{\Omega}(gm)$

$$\frac{d\mathbf{S}}{dt} = \mathbf{\Omega}(gm) \times \mathbf{S}. \tag{1.6}$$

The angular precession velocity $\mathbf{\Omega}(gm)$ of $\mathbf{S}$ around direction of the field $\mathbf{B}$ from gravitomagnetic or electromagnetic origin is given by [6, 7]

$$\mathbf{\Omega}(gm) = -2\beta^{-1} c^{-1} G^{\frac{1}{2}} \mathbf{B}, \tag{1.7}$$

where the precession frequency $\nu(gm)$ is given by $\Omega(gm) = 2\pi\nu(gm)$.

As a first example, the gravitomagnetic precession of the angular momentum $\mathbf{S}_m$ of the circular torus with total mass $m_m$ in the (gravito)magnetic field of the star will be considered. According to (1.6), an angular precession velocity $\mathbf{\Omega}(gm)$ of the component $\mathbf{S}_m \sin\delta_m$ ($\delta_m$ is the angle between the directions of $\mathbf{S}$ and $\mathbf{S}_m$), perpendicular to the angular momentum $\mathbf{S}$ of the star, occurs around $\mathbf{S}$. $\mathbf{\Omega}(gm)$ can directly be calculated from (1.7). Substitution of the equatorial value of the gravitomagnetic field $\mathbf{B}_{eq}(gm) = -R^{-3}\mathbf{M}(gm)$ into (1.7) leads to $\mathbf{\Omega}(gm) = \mathbf{\Omega}_{LT}$, the so-called Lense-Thirring angular precession velocity. When $\delta_m$ is small, $\mathbf{B}_{eq}(gm)$ is approximately constant ($\mathbf{S}_m \sin\delta_m$ reduces to zero for $\delta_m = 0$, so that precession only occurs for $\delta_m > 0$). Furthermore, when the gravitomagnetic dipole moment $\mathbf{M}(gm)$ located at the centre of the star is ideal, combination of (1.1) and (1.7) yields for $\mathbf{\Omega}_{LT}$

$$\mathbf{\Omega}_{LT} = -c^{-2} G R^{-3} \mathbf{S}, \text{ or } \nu_{LT} = -\tfrac{2}{5} c^{-2} G m_s \nu_s r_s^2 R^{-3}. \tag{1.8}$$

The Lense-Thirring frequency $\nu_{LT}$ on the right hand side of (1.8) is obtained by substitution of (1.2) into (1.8a).

For $\delta_m = 90°$ the gravitomagnetic field **B** is no constant and **B** has to be integrated over the whole orbit. An averaged result for the Lense-Thirring frequency $\overline{v_{LT}}$ is then obtained for an exactly circular orbit (although in general relativity orbits need no longer to be closed, a circular orbit remains possible)

$$\overline{v_{LT}} = \tfrac{4}{5} c^{-2} G m_s v_s r_s^2 R^{-3}. \tag{1.9}$$

See the original papers [11] for a discussion of this standard expression of the Lense-Thirring precession. Note that in the derivation of both (1.8) and (1.9) from (1.7) gravitomagnetic fields **B** have been inserted. In that case it makes no difference, whether the field **B** is identified as a common magnetic induction field (we do so) or not. If data are available, predicted values for $v_{LT}$ from (1.8b) for a number of stars are included in the tables below.

A new situation occurs, when an electromagnetic field **B** = **B**(em) is substituted into (1.7). The following different gravitomagnetic precession frequencies can then be distinguished (The adjective "gravitomagnetic" has been added, since (1.7) describes the interaction between some angular momentum and a magnetic field **B**; the field **B**, however, is from electromagnetic origin in this case). Substitution of the field **B**(em) from the torus with total electric charge $Q_o$, acting on the mass current with total mass $m_m$, into (1.7) yields a precession frequency $v_{mo}$. The following sequence with respect to the subscripts will be used throughout this work: the first subscript m in $v_{mo}$ stems from *middle* and the last subscript o from *outer*. The torus with mass $m_m$ in the *middle* experiences the action from the *outer* torus with charge $Q_o$. Likewise, substitution of the field **B**(em) from charge $Q_i$ acting on mass $m_m$ into (1.7) yields a frequency $v_{mi}$. In addition, substitution of the field **B**(em) from charge $Q_o$ acting on mass $m_i$ yields a frequency $v_{io}$. Furthermore, substitution of the field **B**(em) from charge $Q_i$ acting on mass $m_o$ yields a frequency $v_{oi}$. It is stressed, that the precession frequencies $v_{io}$, $v_{mo}$ and $v_{mi}$, and $v_{oi}$ are in fact coupled to the angular momenta of mass $m_i$, $m_m$ and $m_o$, respectively. Note that more than four precession frequencies are obtained, when Lense-Thirring frequencies $v_{LT}$, e.g., from (1.8) are included.

Instead of the gravitomagnetic precession, electromagnetic precession of a torus with a total charge $Q$ and angular momentum **S**, subjected to a magnetic field **B** = **B**(em) is also possible. The angular precession velocity **Ω**(em) of **S** around the direction of the field **B** is given by

$$\mathbf{\Omega}(em) = -\tfrac{1}{2} c^{-1} m^{-1} Q \, \mathbf{B}, \tag{1.10}$$

where $\Omega(em) = 2\pi v(em)$. $Q$ can be equal to $Q_i$ and $m$ to $m_i$, for example. The following combinations will be considered: $Q_i$ and **B**($Q_o$), and $Q_o$ and **B**($Q_i$), resulting in two additional electromagnetic precession frequencies. For the ratio of $v(gm)$ to $v(em)$ follows from combination of (1.7) and (1.10)

$$\Omega(em)/\Omega(gm) = \tfrac{1}{4} \beta (G^{1/2} m)^{-1} Q. \tag{1.11}$$

Thus, apart from $\beta$ the ratio of the frequencies depends on $(G^{1/2} m_i)^{-1} Q_i$ and $(G^{1/2} m_o)^{-1} Q_o$, respectively. We shall discuss $v_{io}(em)$ and $v_{oi}(em)$ for white dwarfs in section 4.

In section 2 we will first consider the high frequency quasi-periodic oscillations (QPOs) and in sections 3 and 4 the low frequency QPOs, due to $Q_o$ and $Q_i$, respectively. A parameter $\beta^*$, describing the total the magnetic field at the pole of the star, $\mathbf{B}_p(tot)$, will be deduced in section 5. A summary of the most important theoretical results will be given in section 6. In sections 7, 8 and 9 observations from four pulsars, a black hole and a white dwarf will be given and discussed, respectively. In section 10 a discussion of the results is given and conclusions are drawn. In section 11 a general summary is given.

## 2. HIGH FREQUENCY QUASI-PERIODIC OSCILLATIONS

First, the calculation of the high frequency QPO $v_i$ will be given. We consider a circular torus containing a total charge $Q_o$, lying in an $x$-$y$ plane at distance $r_o$ from the origin $O$, as shown in figure 1. A radius vector $\mathbf{r}_i$ from $O$ to a point charge $dQ_i$ at field point $F$ is fixed by the spherical coordinates $r_i$, $\theta$ and $\varphi = 0$. The absolute value of the position vector $\mathbf{r}$ from $F$ to a point charge $dQ_o$ in the torus with charge $Q_o$ is then given by

$$r = r_o(1 + x^2 - 2x\sin\theta\cos\varphi)^{1/2}, \qquad (2.1)$$

where $x$ is defined by $x \equiv r_i/r_o$.

Using Coulomb's law, the component of the Coulomb force $d\mathbf{F}_{io}$ in the direction of $\mathbf{r}_i$ between the point charge $dQ_o$ and point charge $dQ_i$ at field point $F$ can then be calculated to be

$$d\mathbf{F}_{io} = \frac{dQ_o dQ_i}{r_o^2} \frac{x - \sin\theta\cos\varphi}{(1 + x^2 - 2x\sin\theta\cos\varphi)^{3/2}} \mathbf{i}_i, \qquad (2.2)$$

where the unit vector $\mathbf{i}_i$ is given by $\mathbf{i}_i = \mathbf{r}_i/|\mathbf{r}_i|$. The following relation applies, when the charge distribution in the torus with total charge $Q_o$ is homogeneous

$$dQ_o = \frac{Q_o}{2\pi} d\varphi. \qquad (2.3)$$

Substitution of the expression for $dQ_o$ from (2.3) into (2.2), followed by integration of $\varphi$ from 0 to $\pi$ yields the following expression for the total Coulomb force $\mathbf{F}_{io}$ from the torus with charge $Q_o$ at field point $F$

$$\mathbf{F}_{io} = \frac{Q_o dQ_i}{\pi r_o^2} \int_0^\pi \frac{x - \sin\theta\cos\varphi}{(1 + x^2 - 2x\sin\theta\cos\varphi)^{3/2}} d\varphi\, \mathbf{i}_i = -f \frac{Q_o dQ_i}{r_o^2} \mathbf{i}_i. \qquad (2.4)$$

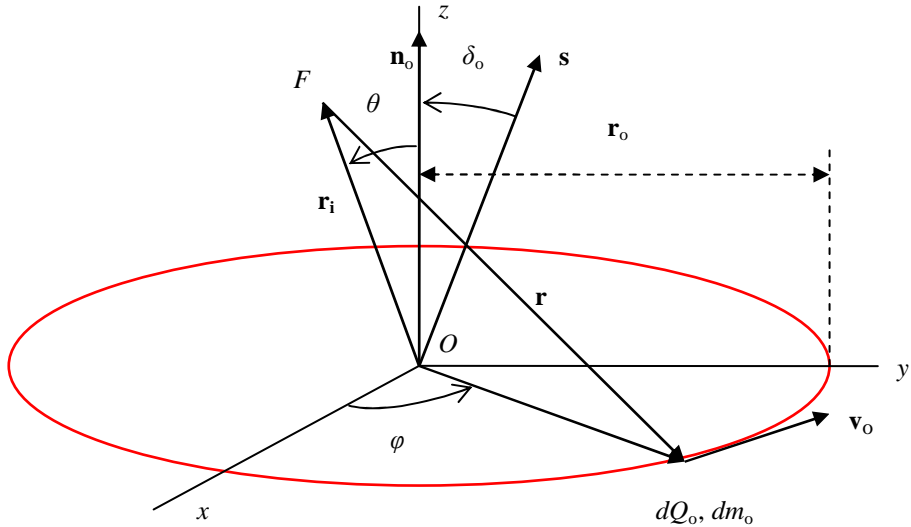

Figure 1. Spherical coordinates $r_i$, $\theta$ and $\varphi = 0$ of the field point $F$ relative to the origin $O$ and the spherical coordinates $r_o$, $\theta = 90°$ and $\varphi = \varphi$ of the point charge $dQ_o$ and point mass $dm_o$, both lying in the $x$-$y$ plane. The angle between the unit vector $\mathbf{s}$, defined as $\mathbf{s} \equiv \mathbf{\Omega}_s/\Omega_s$, and the unit vector $\mathbf{n}_o$ in the direction of the rotation axis of the torus with total charge $Q_o$ has been denoted by $\delta_o$.

Since the charge distribution in the circular torus is symmetric about $\varphi = 0$, the result of the integration for $\mathbf{F}_{io}$ yields no $y$ component for $\varphi = 0$. The function $f$ in (2.4) is defined by

$$f \equiv \frac{-1}{\pi x(1+x^2+2x\sin\theta)^{1/2}} \left\{ K(k) - \frac{(1-x^2)E(k)}{1+x^2-2x\sin\theta} \right\}. \tag{2.5}$$

The integral of (2.4) has been calculated by using complete elliptic integrals of the first kind and second kind, $K(k)$ and $E(k)$, respectively. See for the properties of these integrals, e.g., ref. [12, § 2.57, § 8.11–§ 8.12]. The modulus $k$ of the elliptic integrals is given by

$$k^2 = \frac{4x\sin\theta}{1+x^2+2x\sin\theta}. \tag{2.6}$$

We will treat some properties of this remarkable function $f$ below.

If the star bears a point charge $Q_s$, a point charge $dQ_i$ in the torus with total charge $Q_i$ is subjected to the following Coulomb force $\mathbf{F}_{is}$

$$\mathbf{F}_{is} = \frac{Q_s dQ_i}{r_i^2} \mathbf{i}_i, \tag{2.7}$$

where $\mathbf{i}_i$ is again given by $\mathbf{i}_i = \mathbf{r}_i/|\mathbf{r}_i|$. For a sphere with homogeneous charge density and total charge $Q_s$, series expansion for the total Coulomb force suggests that relation (2.7) remains valid for $r_i$ slightly larger than $r_s$. If equilibrium between the Coulomb forces $\mathbf{F}_{io}$ and $\mathbf{F}_{is}$ exists, combination of (2.4) and (2.7) yields the following relation

$$Q_s = x^2 f Q_o. \tag{2.8}$$

One can also say that the electric fields $\mathbf{F}_{io}/dQ_i$ due to $Q_o$ (see (2.4)) and $\mathbf{F}_{is}/dQ_i$ due to $Q_s$ (see (2.7)) compensate each other. Then, the resulting electric field at field point $F$ is zero. Furthermore, it is noticed that equilibrium is only possible for a positive value of the function $f$. As an illustration, we calculate the value of $f$ in some cases.

It appears that the value $\theta = 90°$ is of particular interest. In that case, the modulus $k$ from (2.6) reduces to $k^2 = 4x/(1+x)^2$ and the following simplified expression for $f$ can be obtained from (2.5)

$$f = \frac{-1}{\pi x}\left\{\frac{K(k)}{1+x} - \frac{E(k)}{1-x}\right\} = \frac{-2}{\pi x}\left\{K(x) - \frac{E(x)}{1-x^2}\right\} \equiv f(x). \tag{2.9}$$

In deriving the right hand side of (2.9), use has been made of the following properties of the complete elliptic integrals: $K(k) = (1+x)K(x)$ and $E(k) = \{2E(x)/(1+x)\} - (1-x)K(x)$ (see, e.g., ref. [12, § 8.12]).

A second case follows from (2.5), when $f \equiv f_0 = 0$

$$K(k_0) = \frac{(1-x_0^2)E(k_0)}{1+x_0^2-2x_0\sin\theta_0}. \tag{2.10}$$

An additional relation between $k_0$, $x_0$ and $\theta_0$ is then given by (2.6). When a value for $k_0$ is chosen $K(k_0)$ and $E(k_0)$ can be found. By combining (2.6) and (2.10) the values for $x_0$ and $\theta_0$ can be calculated. Further, it is noticed that $k_0$ and $k$ can be written as

$$k_0 = \sin\alpha_0 \text{ and } k = \sin\alpha, \tag{2.11}$$

where $\alpha_0$ and $\alpha$ are modular angles.

In table 1 a number of parameters occurring in (2.5), (2.6), (2.9), (2.10) and (2.11) are given. The procedure to calculate the various parameters has been as follows. First, a value of $\alpha_0$ is chosen and $k_0$ is calculated from (2.11a). When $k_0$ is known, $K(k_0)$ and $E(k_0)$ can be calculated from their respective series expansions (see, e.g., [12, § 8.11]). Then, $x_0$ and $\theta_0$ can be calculated from (2.6) and (2.10). Subsequently, choosing $\theta = 90°$ and $x = x_0$, the values of $k$ and $\alpha$ can be calculated from (2.6) and (2.11b), respectively. When $k$ is known, the complete elliptic integrals $K(k)$ and $E(k)$ can also be calculated from their respective series expansions (see again, e.g., [12, § 8.11]). In addition, the value for $f$ can be calculated from (2.9). It is noticed that for $x = x_0$ the charge system is only stable, if $\theta$ is lying in the interval $\theta_0 < \theta \leq 90°$, corresponding to the interval $f_0 < f \leq f(x)$ for $f$.

A third case follows from (2.6), when $\theta = 0°$. A value $k = 0$ is then obtained. Introduction of the quantities $\theta = 0°$ and $k = 0$ into (2.5) yields for $f$

$$f = -\frac{x}{(1+x^2)^{3/2}}. \qquad (2.12)$$

Note that $f$ is negative for a positive value of $x$, so that no equilibrium can exist for the chosen charge configuration when only Coulomb forces are present. Thus, the factor $f$ of (2.5) is an important parameter with respect to the stability of the system.

In order to illustrate the property of sign change of function $f$, the following example will be considered. The plane of the torus with total charge $Q_i$ and the plane of the torus with total charge $Q_o$ are supposed to have the $y$ axis in common. The charge elements $dQ_i$ at the field point $F$ in the $x$-$z$ plane with the spherical coordinates $r_i$, $\theta = \theta$ and $\varphi = 0$ (see figure 2) and at the point with coordinates $r_i$, $\theta = 180° - \theta$ and $\varphi = 180°$ are less firmly bound (or in a non-bound state when $\theta < \theta_0$). However, the charges $dQ_i$ at the coordinates $r_i$, $\theta = 90°$ and $\varphi = 90°$ and $r_i$, $\theta = 90°$ and $\varphi = 270°$ are in a bound state. As a result, the current flows in the tori may be interrupted twice per revolution around the star, so that their frequencies may become manifest. From the results in table 1 follows that values for $\theta_0$ are only a little bit smaller than 90° when $x$ approach unity value. Thus, the stability range for $\theta$ is rather limited for large values of $x$. In order to obtain a stable situation between the tori with charges $Q_i$ and $Q_o$, respectively, the angle $\Delta = 90° - \theta$ between the unit vectors $\mathbf{n}_i$ and $\mathbf{n}_o$ (see figure 2) has to be small. The sign change of the function $f$ might thus explain the manifestation of the high frequency QPOs $\nu_i$ and $\nu_o$. The occurrence of the high frequency QPO $\nu_m$, however, is not explained by this mechanism. For simplicity reasons, it will be assumed in this work that the orbital frequencies $\nu_i$, $\nu_o$ and $\nu_m$ represent the number of complete revolutions per second around the star.

For the case considered in figure 2, the value of $f$ can be calculated at the field point $F$ of any point charge $dQ_i$. For example, for $\theta = \theta_0$ at the field points $F(r_i, \theta_0, \varphi = 0)$ and $F(r_i, \theta_0, \varphi = 180°)$ $f = f_0 = 0$ and for both field points $F(r_i, \theta = 90°, \varphi = 90°)$ and $F(r_i, \theta = 90°, \varphi = 270°)$ $f = f(x)$ from (2.9). Likewise, other values for $f$ can be calculated from (2.5). Subsequently, the average value of $f$ over the whole orbit with total charge $Q_i$ could be calculated. In our calculations we will approximate the average value of $f$ over the whole orbit, $\bar{f}(x)$, by $\bar{f}(x) = \frac{1}{2}\{f_0 + f(x)\} = \frac{1}{2}\{0 + f(x)\} = \frac{1}{2}f(x)$.

In deriving (2.8), only Coulomb forces have been taken into account. When the gravitation law of Newton and the centrifugal force are also included, a more general description is obtained. When all these forces between a point mass $dm_i$ with charge $dQ_i$ in the inner torus, a point charge $Q_s$ in the star and a charge $Q_o$ in the outer torus are in equilibrium, the following relation can be calculated

$$\frac{dm_i v_i^2}{r_i} = \frac{dm_i G m_s}{r_i^2} - \frac{dQ_i Q_s}{r_i^2} + f \frac{dQ_i Q_o}{r_o^2}, \qquad (2.13)$$

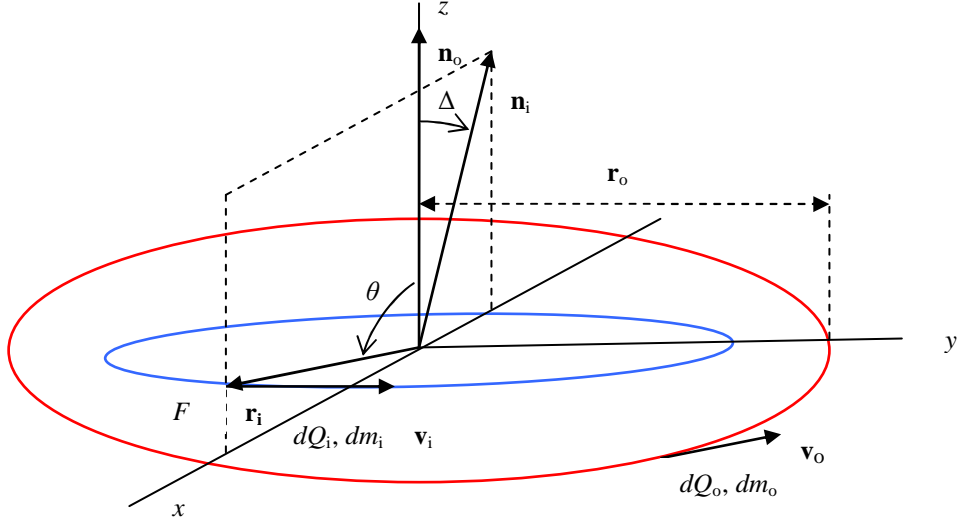

Figure 2. Orbits of the tori with total charge $Q_o$ (red) and $Q_i$ (blue), respectively. The direction of the positive $z$ axis of the coordinate system is chosen along the rotation axis of the torus with charge $Q_o$. The unit vector in this direction is denoted by $\mathbf{n}_o$. The direction of the rotation axis of the torus with charge $Q_i$ with unit vector $\mathbf{n}_i$ is chosen in the $x$-$z$ plane. The angle between $\mathbf{n}_o$ and $\mathbf{n}_i$ is given by $\Delta$. The field point $F$ with spherical coordinates $r_i$, $\theta$ and $\varphi = 0$ is shown (see also text).

where $v_i$ is the velocity of $dm_i$ moving in a circular orbit with radius $r_i$ around the star with mass $m_s$. The charge $Q_o$ with corresponding mass $m_o$ also moves in a circular orbit with radius $r_o$ around the star ($r_i < r_o$). It is noticed that the gravitational attraction between the masses $m_i$ and $m_o$ in the tori and Lorentz forces in the considered system have been neglected. Moreover, no general relativistic effects have been taken into account in the derivation of (2.13). Starting from a Kerr-Newman space-time, Aliev and Galtsov [13] considered the latter effects for the binary system of a charged star and a charged mass moving in a circular orbit around that star. Therefore, equation (2.13) has to be considered as a first order approximation.

For a homogeneous mass and charge distribution in the torus with a total mass $m_i$ and a total charge $Q_i$, respectively, the following relation applies

$$\frac{dQ_i}{dm_i} = \frac{Q_i}{m_i}. \tag{2.14}$$

Then, from (2.13) and (2.14) the following expression for the high frequency $v_i$ for a point mass $dm_i$ with charge $dQ_i$ in the torus with a total charge $Q_i$ can be calculated

$$v_i = \frac{1}{2\pi}\left[\frac{Gm_s}{r_i^3}\left\{1 - \frac{m_s}{m_i}Q_i'\left(Q_s' - x^2 f Q_o'\right)\right\}\right]^{1/2}, \tag{2.15}$$

where $Q_i'$ is defined by the dimensionless quantity $Q_i' \equiv (G^{1/2}m_s)^{-1}Q_i$, $Q_s'$ by $Q_s' \equiv (G^{1/2}m_s)^{-1}Q_s$ and so on. It is to be expected that the factor $m_s/m_i$ on the right hand side of (2.15) is large. In this work it will be assumed that the difference $(Q_s - x^2 f Q_o)$ is small, so that relation (2.8) remains approximately valid. The charge dependent contribution on the right hand side of (2.15) may then be positive or negative. As a result, the orbital frequency $v_i$ may thus be larger or smaller than the corresponding Kepler frequency $(2\pi)^{-1}(Gm_s/r_i^3)^{1/2}$.

Table 1. Calculated values for $k_0$, $x_0$, $\theta_0$, $g_0$ and $f_0$ ($f_0 = 0$), together with the values for $K(k_0)$ and $E(k_0)$ on the upper line. Choosing $\theta = 90º$ and $x = x_0$, one obtains the values for $k$, $K(k)$, $E(k)$, $f(x)$ and $g(x)$. The latter values have been given on the lower line.

| $\alpha_0$ | $k_0$ | $K(k_0)$ | $E(k_0)$ | $x_0$ | $\theta_0$ (º) | $f_0$ | $g_0$ |
| $\alpha$ | $k$ | $K(k)$ | $E(k)$ | $x = x_0$ | $\theta$ (º) | $f(x)$ | $g(x)$ |
|---|---|---|---|---|---|---|---|
| 89.50000 | 0.9999619 | 6.1278 | 1.0002 | 0.99907 | 89.001 | 0 | 1.952 |
| 89.97325 | 0.9999999 | 9.0565 | 1.0000 | 0.99907 | 90 | 340.2 | 342.8 |
| 89.20000 | 0.9999025 | 5.6579 | 1.0005 | 0.99780 | 88.41 | 0 | 1.803 |
| 89.93687 | 0.9999994 | 8.1967 | 1.0000 | 0.99780 | 90 | 143.6 | 145.9 |
| 89.00000 | 0.9998477 | 5.4349 | 1.0008 | 0.99670 | 88.01 | 0 | 1.733 |
| 89.90526 | 0.9999986 | 7.7911 | 1.0000 | 0.99670 | 90 | 95.48 | 97.65 |
| 88.0000 | 0.999391 | 4.7427 | 1.0026 | 0.98856 | 86.06 | 0 | 1.519 |
| 89.6703 | 0.999983 | 6.5441 | 1.0001 | 0.98856 | 90 | 27.08 | 28.87 |
| 87.0000 | 0.998630 | 4.3387 | 1.0053 | 0.97669 | 84.16 | 0 | 1.399 |
| 89.3244 | 0.999930 | 5.8269 | 1.0004 | 0.97669 | 90 | 13.03 | 14.60 |
| 86.0000 | 0.997564 | 4.0528 | 1.0086 | 0.96181 | 82.34 | 0 | 1.318 |
| 88.8847 | 0.999811 | 5.3259 | 1.0009 | 0.96181 | 90 | 7.776 | 9.208 |
| 85.0000 | 0.996195 | 3.8317 | 1.0127 | 0.94449 | 80.58 | 0 | 1.259 |
| 88.3640 | 0.999592 | 4.9432 | 1.0018 | 0.94449 | 90 | 5.225 | 6.554 |
| 84.0000 | 0.994522 | 3.6519 | 1.0172 | 0.92519 | 78.92 | 0 | 1.213 |
| 87.7730 | 0.999245 | 4.6355 | 1.0031 | 0.92519 | 90 | 3.785 | 5.035 |
| 83.0000 | 0.992546 | 3.5004 | 1.0223 | 0.90433 | 77.33 | 0 | 1.177 |
| 87.1205 | 0.998737 | 4.3794 | 1.0049 | 0.90433 | 90 | 2.888 | 4.076 |
| 82.0000 | 0.990268 | 3.3699 | 1.0278 | 0.88228 | 75.83 | 0 | 1.148 |
| 86.4144 | 0.998042 | 4.1613 | 1.0072 | 0.88228 | 90 | 2.289 | 3.427 |
| 81.0000 | 0.987688 | 3.2553 | 1.0338 | 0.85934 | 74.41 | 0 | 1.125 |
| 85.6615 | 0.997135 | 3.9722 | 1.0099 | 0.85934 | 90 | 1.868 | 2.966 |
| 80.0000 | 0.984808 | 3.1534 | 1.0401 | 0.83578 | 73.07 | 0 | 1.105 |
| 84.8679 | 0.995991 | 3.8060 | 1.0132 | 0.83578 | 90 | 1.560 | 2.624 |
| 79.0000 | 0.981627 | 3.0617 | 1.0468 | 0.81183 | 71.81 | 0 | 1.0893 |
| 84.0387 | 0.994592 | 3.6582 | 1.0171 | 0.81183 | 90 | 1.328 | 2.363 |
| 78.0000 | 0.978148 | 2.9786 | 1.0538 | 0.78767 | 70.63 | 0 | 1.0758 |
| 83.1786 | 0.992921 | 3.5258 | 1.0214 | 0.78767 | 90 | 1.147 | 2.159 |
| 77.0000 | 0.974370 | 2.9026 | 1.0611 | 0.76347 | 69.52 | 0 | 1.0645 |
| 82.2917 | 0.990964 | 3.4061 | 1.0262 | 0.76347 | 90 | 1.0035 | 1.996 |
| 76.0000 | 0.970296 | 2.8327 | 1.0686 | 0.73935 | 68.48 | 0 | 1.0549 |
| 81.3815 | 0.988708 | 3.2974 | 1.0315 | 0.73935 | 90 | 0.8876 | 1.863 |
| 75.0000 | 0.965926 | 2.7681 | 1.0764 | 0.71543 | 67.51 | 0 | 1.0468 |
| 80.4510 | 0.986144 | 3.1980 | 1.0372 | 0.71543 | 90 | 0.7922 | 1.754 |
| 74.0000 | 0.961262 | 2.7081 | 1.0844 | 0.69178 | 66.60 | 0 | 1.0399 |
| 79.5029 | 0.983264 | 3.1067 | 1.0434 | 0.69178 | 90 | 0.7127 | 1.662 |
| 73.0000 | 0.956305 | 2.6521 | 1.0927 | 0.66849 | 65.75 | 0 | 1.0341 |
| 78.5396 | 0.980062 | 3.0225 | 1.0500 | 0.66849 | 90 | 0.6455 | 1.585 |
| 72.0000 | 0.951057 | 2.5998 | 1.1011 | 0.64559 | 64.96 | 0 | 1.0291 |
| 77.5630 | 0.976533 | 2.9445 | 1.0569 | 0.64559 | 90 | 0.5882 | 1.519 |
| 71.0000 | 0.945519 | 2.5507 | 1.1096 | 0.62315 | 64.22 | 0 | 1.0249 |
| 76.5749 | 0.972674 | 2.8722 | 1.0642 | 0.62315 | 90 | 0.5387 | 1.462 |
| 70.0000 | 0.939693 | 2.5045 | 1.1184 | 0.60118 | 63.52 | 0 | 1.0212 |
| 75.5768 | 0.968483 | 2.8047 | 1.0719 | 0.60118 | 90 | 0.4956 | 1.413 |
| 69.000 | 0.93358 | 2.4610 | 1.1272 | 0.57970 | 62.88 | 0 | 1.0181 |
| 74.570 | 0.96396 | 2.7417 | 1.0798 | 0.57970 | 90 | 0.4577 | 1.3703 |
| 68.000 | 0.92718 | 2.4198 | 1.1362 | 0.55875 | 62.28 | 0 | 1.0155 |
| 73.556 | 0.95910 | 2.6827 | 1.0881 | 0.55875 | 90 | 0.4243 | 1.3327 |
| 67.000 | 0.92050 | 2.3809 | 1.1453 | 0.53831 | 61.72 | 0 | 1.0132 |
| 72.535 | 0.95390 | 2.6274 | 1.0965 | 0.53831 | 90 | 0.3945 | 1.2997 |
| 66.000 | 0.91355 | 2.3439 | 1.1545 | 0.51840 | 61.20 | 0 | 1.0113 |
| 71.508 | 0.94837 | 2.5753 | 1.1053 | 0.51840 | 90 | 0.3678 | 1.2704 |

| $\alpha_0$ | $k_0$ | $K(k_0)$ | $E(k_0)$ | $x_0$ | $\theta_0$ (°) | $f_0 = 0$ | $g_0$ |
| $\alpha$ | $k$ | $K(k)$ | $E(k)$ | $x = x_0$ | $\theta$ (°) | $f(x)$ | $g(x)$ |
|---|---|---|---|---|---|---|---|
| 65.000 | 0.90631 | 2.3088 | 1.1638 | 0.49903 | 60.71 | 0 | 1.0096 |
| 70.476 | 0.94250 | 2.5262 | 1.1142 | 0.49903 | 90 | 0.3437 | 1.2444 |
| 64.000 | 0.89879 | 2.2754 | 1.1732 | 0.48017 | 60.26 | 0 | 1.0082 |
| 69.440 | 0.93630 | 2.4798 | 1.1233 | 0.48017 | 90 | 0.3219 | 1.2211 |
| 63.000 | 0.89101 | 2.2435 | 1.1826 | 0.46185 | 59.84 | 0 | 1.0070 |
| 68.399 | 0.92977 | 2.4360 | 1.1326 | 0.46185 | 90 | 0.3021 | 1.2004 |
| 62.000 | 0.88295 | 2.2132 | 1.1920 | 0.44403 | 59.45 | 0 | 1.0059 |
| 67.356 | 0.92291 | 2.3945 | 1.1421 | 0.44403 | 90 | 0.2839 | 1.1817 |
| 61.000 | 0.87462 | 2.1842 | 1.2015 | 0.42673 | 59.08 | 0 | 1.0050 |
| 66.309 | 0.91572 | 2.3551 | 1.1517 | 0.42673 | 90 | 0.2673 | 1.1649 |
| 60.000 | 0.86603 | 2.1565 | 1.2111 | 0.40992 | 58.74 | 0 | 1.0043 |
| 65.259 | 0.90821 | 2.3177 | 1.1614 | 0.40992 | 90 | 0.2519 | 1.1498 |

Secondly, the calculation of the high frequency $v_o$ will be given. We now consider a circular torus containing a total charge $Q_i$, lying in another $x$-$y$ plane at distance $r_i$ from the origin $O$ (compare with figure 1). A radius vector $\mathbf{r}_o$ from $O$ to field point $F$ is fixed by the spherical coordinates $r_o$, $\theta$ and $\varphi = 0$. The absolute value of the position vector $\mathbf{r}$ from $F$ to a point charge $dQ_i$ in the torus with charge $Q_i$ is then also given by (2.1).

Using Coulomb's law, the component of the Coulomb force $d\mathbf{F}_{oi}$ in the direction of $\mathbf{r}_o$ at the field point $F$ from the point charge $dQ_i$ can then be calculated to be

$$d\mathbf{F}_{oi} = \frac{dQ_o dQ_i}{r_o^2} \frac{1 - x\sin\theta\cos\varphi}{(1 + x^2 - 2x\sin\theta\cos\varphi)^{3/2}} \mathbf{i}_o, \qquad (2.16)$$

where the unit vector $\mathbf{i}_o$ is given by $\mathbf{i}_o = \mathbf{r}_o/|\mathbf{r}_o|$. Assuming that the charge distribution in the torus with total charge $Q_i$ is also homogeneous, an expression for $dQ_i$, analogous to $dQ_o$ from (2.3), can be substituted into (2.16). Integration of $\varphi$ from 0 to $\pi$ of the resulting expression yields the following equation for the total Coulomb force $\mathbf{F}_{oi}$ from the torus with charge $Q_i$ at field point $F$

$$\mathbf{F}_{oi} = \frac{Q_i dQ_o}{\pi r_o^2} \int_0^\pi \frac{1 - x\sin\theta\cos\varphi}{(1 + x^2 - 2x\sin\theta\cos\varphi)^{3/2}} d\varphi \, \mathbf{i}_o = g \frac{Q_i dQ_o}{r_o^2} \mathbf{i}_o. \qquad (2.17)$$

Since the charge distribution in the circular torus is symmetric about $\varphi = 0$, the result of the integration for $\mathbf{F}_{oi}$ contains no $y$ component for $\varphi = 0$. The function $g$ in (2.17) is defined by

$$g \equiv \frac{1}{\pi(1 + x^2 + 2x\sin\theta)^{1/2}} \left\{ K(k) + \frac{(1 - x^2)E(k)}{1 + x^2 - 2x\sin\theta} \right\}. \qquad (2.18)$$

The integral of (2.17) has also been calculated by using the complete elliptic integrals of the first kind and second kind, $K(k)$ and $E(k)$, respectively. See for the properties of these integrals, e.g., [12, § 2.57, § 8.11– § 8.12]. The modulus $k$ of the elliptic integrals is again given by (2.6).

In addition, a point charge $dQ_o$ in the torus with total charge $Q_o$ is subjected to the following Coulomb force $\mathbf{F}_{os}$ from the total charge $Q_s$ of the star, analogously to the Coulomb force $\mathbf{F}_{is}$ of (2.7),

$$\mathbf{F}_{os} = \frac{Q_s dQ_o}{r_o^2} \mathbf{i}_o, \qquad (2.19)$$

where $\mathbf{i}_o$ is again given by $\mathbf{i}_o = \mathbf{r}_o/|\mathbf{r}_o|$. If equilibrium between the Coulomb forces $\mathbf{F}_{oi}$ and $\mathbf{F}_{os}$ exists, combination of (2.17) and (2.19) yields the following relation

$$Q_s = -g\, Q_i. \tag{2.20}$$

One can again say that the electric fields $\mathbf{F}_{oi}/dQ_o$ due to $Q_i$ (see (2.17)) and $\mathbf{F}_{os}/dQ_o$ due to $Q_s$ (see (2.19)), respectively, then compensate each other. Thus, the resulting electric field at this field point $F$ is zero, too.

The general expression $g$ of (2.18) will now be evaluated in some cases. For $\theta = 90°$, the modulus $k$ of (2.6) again reduces to $k^2 = 4x/(1+x)^2$, whereas $g$ of (2.18) reduces to

$$g = \frac{1}{\pi}\left\{\frac{K(k)}{1+x} + \frac{E(k)}{1-x}\right\} = \frac{2}{\pi}\left\{\frac{E(x)}{1-x^2}\right\} \equiv g(x). \tag{2.21}$$

In deriving (2.21), use has again been made of the relations: $K(k) = (1+x)K(x)$ and $E(k) = \{2E(x)/(1+x)\} - (1-x)K(x)$ (see, e.g., [12, § 8.12]). Taking $x = x_0$, $\theta_0$, $k_0$ and $k$ from table 1, $g_0$ can be calculated from (2.18) and $g(x)$ from (2.21). Results for $g_0$ and $g(x)$ have been added to table 1 for a number of cases.

When $\theta = 0°$, $k = 0$ follows from (2.6). Introduction of the quantities $\theta = 0°$, $k = 0$ into (2.18) then yields for $g$

$$g = \frac{1}{(1+x^2)^{3/2}}. \tag{2.22}$$

Relations (2.21) and (2.22) show that function $g$ is positive in both cases.

Analogously to the method followed for $\bar{f}(\bar{x})$, from (2.18) an average value for $g$ can be calculated. For the case considered in figure 2, the value of $g$ could be calculated at the field point $F$ for any point charge $dQ_o$ in the torus with total charge $Q_o$. Subsequently, the average value of $g$ over the whole orbit could be deduced. Utilizing (2.21) in our calculations, we will approximate $\bar{g}(\bar{x})$ by $\bar{g}(\bar{x}) = \frac{1}{2}\{g_0 + g(x)\}$.

In deriving (2.20), only Coulomb forces have been taken into account. When the gravitation law of Newton and the centrifugal force are also included, a more general description is obtained. When all these forces between a point mass $dm_o$ with charge $dQ_o$ in a torus, a charge $Q_s$ in the star and a charge $Q_i$ in another torus are in equilibrium, the following relation, analogous to (2.13), can be calculated

$$\frac{dm_o v_o^2}{r_o} = \frac{dm_o G m_s}{r_o^2} - \frac{dQ_o Q_s}{r_o^2} - g\frac{dQ_o Q_i}{r_o^2}, \tag{2.23}$$

where $v_o$ is the velocity of $dm_o$ moving in a circular orbit with radius $r_o$ around the star with mass $m_s$ (see figures 1 and 2). The charge $Q_i$ with corresponding mass $m_i$ also moves in a circular orbit of radius $r_i$ around the star ($r_i < r_o$). In deducing (2.23), similar approximations have been applied as in the derivation of (2.13). Moreover, it is assumed that the velocities $v_i$ in (2.13) and $v_o$ in (2.23) are non-relativistic (In the relativistic case, equations (2.13) and (2.23) are no longer valid).

In order to calculate the high frequency $v_o$, one may use a relation analogous to (2.14), assuming a homogeneous mass and charge distribution in the torus with a total mass $m_o$ and a total charge $Q_o$, respectively. Combination of this relation with (2.23) then yields the following expression for $v_o$ for a point mass $dm_o$ with charge $dQ_o$ in the torus with a total charge $Q_o$

$$v_o = \frac{1}{2\pi}\left[\frac{Gm_s}{r_o^3}\left\{1-\frac{m_s}{m_o}Q_o'\left(Q_s'+gQ_i'\right)\right\}\right]^{1/2}, \qquad (2.24)$$

where $Q_o'$ is defined by the dimensionless quantity $Q_o' \equiv (G^{1/2}m_s)^{-1}Q_o$ and so on. It is to be expected that the factor $m_s/m_o$ on the right hand side of (2.24) is large. In this work it will be assumed that the sum $(Q_s + g Q_i)$ is small, so that relation (2.20) remains approximately valid. In principle the charge dependent contribution on the right hand side of (2.24) may thus be positive or negative. The orbital frequency $v_o$ may thus be larger or smaller than the corresponding Kepler frequency $(2\pi)^{-1}(Gm_s/r_o^3)^{1/2}$.

Using (2.8) and (2.20), it is possible to calculate the total charge $Q_{tot}$ of the system One obtains

$$Q_{tot} = Q_s + Q_i + Q_o = (1-\frac{1}{g}+\frac{1}{x^2 f})Q_s. \qquad (2.25)$$

This relation may be helpful to understand the loading mechanism of a star. Suppose, for example, that the quantity $x$ will increase as a result of expansion of the inner torus during nuclear burning at the surface of the star, or as a result of compression of the outer torus, due to accretion. Then, it follows from (2.25) and data in table 1, that the quantity $\{1 - 1/g + 1/(x^2 f)\}$ becomes less positive. Assuming that $Q_{tot}$ is constant and that the charge $Q_s$ is positive, the amount of positive charge $Q_s$ will increase and, e.g., positive charge must flow to the star. As another consequence, the Lorentz force will generate a toroidal current in the star, when the positive charge enters the star at the equator.

In addition, combination of (2.8) and (2.20) leads to the following relation between $Q_i$ and $Q_o$

$$Q_i = -\frac{x^2 f}{g}Q_o. \qquad (2.26)$$

If the planes of the tori of the charges $Q_i$ and $Q_o$ coincide, it can be shown that for $x \to 0$ $Q_i \to 0$ and for $x \to 1$ $Q_i \to -Q_o$.

Finally, the orbital frequency $v_m$ for a point mass $dm_m$ in a circular torus containing a homogeneous electrically neutral mass distribution with total mass $m_m$ can be shown to be (compare with, e.g., Aliev and Galtsov [13])

$$v_m = \frac{1}{2\pi}\left(\frac{Gm_s}{r_m^3}\right)^{1/2}\frac{1}{1+\frac{S}{c^2 m_s}\left(\frac{Gm_s}{r_m^3}\right)^{1/2}} = \frac{1}{2\pi}\left(\frac{Gm_s}{r_m^3}\right)^{1/2}f_S \approx \frac{1}{2\pi}\left(\frac{Gm_s}{r_K^3}\right)^{1/2} = v_K, \qquad (2.27)$$

where $r_m$ is the radius of the torus with mass $m_m$ and $v_K$ is the Kepler frequency of the point mass $dm_m$ moving around the star ($r_K$ is the corresponding Kepler radius). The result applies for prograde motion of $dm_m$ in the equatorial plane around the star (Thus, the directions of the angular momentum $\mathbf{S}_m$ of the torus with mass $m_m$ and the angular momentum $\mathbf{S}$ of the star are taken parallel). In the model proposed by Stella, Vietri and Morsink [4] the QPO with the highest frequency, the so-called *upper* kHz frequency, $v_u$, was identified with our frequency $v_m$ from (2.27). Usually, the relativistic factor $f_S$ in (2.27) depending on the angular momentum $S$ approaches unity value, so that the frequency $v_m$ can be approximated by the Kepler frequency $v_K$. In deriving (2.27), other relativistic and classical interactions with the other tori have been neglected (compare with comment to (2.13) and (2.23)).

It is noticed that the frequencies $\nu_i$, $\nu_o$ and the approximated frequency $\nu_m$ can all be obtained by application of classical mechanics and Coulomb's law only. Moreover, none of these orbital frequencies depend on the rotational frequency $\nu_s$ of the star. In section 5 of this work it will be discussed how the influence of $\nu_s$ may become manifest in another way.

As an aside, the electric fields at the pole and the equator of the star for $\delta_i = \delta_o = 0°$ can easily be calculated from this section. By combining equations analogous to (2.4), (2.7), (2.8), (2.12) and (2.20), the total electric field $\mathbf{E}_p(\text{tot})$ at the north pole of the star, due to the charge $Q_s$ in the star and the charges $Q_i$ and $Q_s$ in the inner and outer torus, respectively, can be deduced. One obtains

$$\mathbf{E}_p(\text{tot}) = \frac{Q_s}{r_s^2}\left[1 + \frac{r_s^2}{r_i^2}\left\{-\frac{1}{g(x)}\frac{r_s/r_i}{\left(1+r_s^2/r_i^2\right)^{3/2}} + \frac{1}{f(x)}\frac{r_s/r_o}{\left(1+r_s^2/r_o^2\right)^{3/2}}\right\}\right]\mathbf{s}, \qquad (2.28)$$

where $x = r_i/r_o$, $\mathbf{s} = \mathbf{\Omega}_s/\Omega_s$ and $f(x)$ and $g(x)$ are given by (2.9) and (2.21), respectively. Relation (2.28) illustrates that charge equilibrium between the star and the charged tori in the equatorial plane (embodied in (2.8) and (2.20)) does not necessarily lead to charge equilibrium along the $\mathbf{s}$ axis. As a result, outflow of charge may take place at the poles (*jets*). However, this subject will be considered in the future.

In addition, by combining equations analogous to (2.4), (2.7), (2.8) and (2.20), the radial electric field $E_{eq}(\text{tot})$ at the equator can be calculated to be

$$E_{eq}(\text{tot}) = \frac{Q_s}{r_s^2}\left[1 + \frac{r_s^2}{r_i^2}\left\{\frac{f(r_s/r_i)}{g(x)} - \frac{f(r_s/r_o)}{f(x)}\right\}\right], \qquad (2.29)$$

where again $x = r_i/r_o$ and $f(r_s/r_i)$ and $f(x)$ can be found from (2.9), and $g(r_s/r_i)$ and $g(x)$ from (2.21). When sufficient parameters are known, values for $E_p(\text{tot})$ and $E_{eq}(\text{tot})$ can be calculated. It is probable that large values of $E_{eq}(\text{tot})$ induce phenomena like sparking, lightning and X-ray emission at the equator of the star.

In sections 3 and 4, we will now discuss the so-called low frequency quasi-periodic oscillations.

## 3. LOW FREQUENCY QUASI-PERIODIC OSCILLATIONS, DUE TO $Q_o$

First, the calculation of the gravitomagnetic, low frequency QPO $\nu_{io}$ will be given. We again consider the circular torus containing a total charge $Q_o$, lying in an $x$-$y$ plane at distance $r_o$ from the origin $O$, as shown in figure 1. A radius vector $\mathbf{r}_i$ from $O$ to field point $F$, where a point charge $dQ_i$ is situated, is fixed by the spherical coordinates $r_i$, $\theta$ and $\varphi = 0$. The current $dQ_o/dt$ from the charge in the torus generates a vector potential in the $y$ direction at field point $F$, $A_\varphi(r_i, \theta)$, of magnitude (compare with derivation of, e.g., Jackson [14])

$$A_\varphi(r_i, \theta) = \frac{1}{c}\frac{dQ_o}{dt}\int_0^{2\pi} \frac{\cos\varphi \, d\varphi}{\left(1 + x^2 - 2x\sin\theta\cos\varphi\right)^{1/2}}, \qquad (3.1)$$

where $x$ is defined by $x \equiv r_i/r_o$. It is noticed that only the $y$ component of the current $dQ_o/dt$ contributes to $A_\varphi(r_i, \theta)$. For symmetry reasons the net $x$ component reduces to zero value. The integral of (3.1) can be calculated by using complete elliptic integrals of the first kind and second kind, $K(k)$ and $E(k)$, respectively. See, e.g., ref. [12] for the properties of these integrals. Moreover, the relation $dQ_o/dt = Q_o\nu_o$ is substituted into (3.1). One obtains

$$A_\varphi(r_i, \theta) = \frac{4Q_o v_o}{c(1+x^2+2x\sin\theta)^{1/2}} \left\{ \frac{(2-k^2)K(k) - 2E(k)}{k^2} \right\}, \tag{3.2}$$

where the modulus $k$ of the elliptic integrals is again given by (2.6).

Introduction of $\theta = 90°$ and the reduced modulus $k^2 = 4x/(1+x)^2$ from (2.6) into (3.2), followed by evaluation of the elliptic integrals $K(k)$ and $E(k)$ (see, e.g., [12, § 8.12]), leads to the following simplified expression for $A_\varphi(r_i, \theta = 90°)$

$$A_\varphi(r_i, \theta=90°) = \frac{4Q_o v_o}{cx}\{K(x) - E(x)\}. \tag{3.3}$$

Likewise, for $\theta \to 0°$ a value $k \to 0$ follows from (2.6). Introduction of the quantities $\theta \to 0°$ and $k \to 0$ into (3.2) then yields for $A_\varphi(r_i, \theta \to 0°)$

$$A_\varphi(r_i, \theta \to 0°) = \frac{\pi Q_o v_o}{c} \frac{x\sin\theta}{(1+x^2+2x\sin\theta)^{3/2}}. \tag{3.4}$$

Equation (3.4) will be used in the calculations of section 5.

From $A_\varphi(r_i, \theta)$ the components of the magnetic induction field

$$\begin{aligned} B_{r_i} &= \frac{1}{r_i \sin\theta} \frac{\partial}{\partial \theta}(\sin\theta A_\varphi) \\ B_\theta &= -\frac{1}{r_i} \frac{\partial}{\partial r_i}(r_i A_\varphi) \\ B_\varphi &= 0 \end{aligned} \tag{3.5}$$

can be calculated, although the results are usually rather complex. We restrict ourselves to the calculation of $B_\theta(r_i, \theta = 90°)$ from (3.3) and (3.5), resulting into

$$B_\theta(r_i, \theta=90°) = -\frac{2\pi Q_o v_o}{cr_o}\left\{\frac{2}{\pi}\frac{E(x)}{(1-x^2)}\right\} = -\frac{2\pi Q_o v_o}{cr_o}g(x). \tag{3.6}$$

In deriving (3.6), use has been made of the properties of complete elliptic integrals (compare with the derivation of (2.21)). Note that the quantity $g(x)$ of (3.6) equals to $g(x)$ from (2.21). The field **B**(em) from charge $Q_o$ acting on the component $\mathbf{S}_i\sin\Delta$ can be calculated from (3.6) ($\mathbf{S}_i$ is the angular momentum of the torus with mass $m_i$ and charge $Q_i$; for the definition of $\Delta$, see figure 2. Precession only occurs for $\Delta > 0$). One obtains in a first approximation

$$\mathbf{B}(\text{em}) \approx \frac{2\pi Q_o v_o}{cr_o}g(x)\cos\Delta\,\mathbf{n}_o \approx \frac{2\pi Q_o v_o}{cr_o}g(x)\cos\delta_i\cos\delta_o\,\mathbf{n}_o, \tag{3.7}$$

where $\mathbf{n}_o$ is the unit vector in the direction of the rotation axis of the torus with charge $Q_o$ and $\delta_o$ is the angle between the direction rotation axis $\mathbf{s} = \mathbf{\Omega}_s/\Omega_s$ of the star and the unit vector $\mathbf{n}_o$. In addition, $\delta_i$ is the angle between the unit vector $\mathbf{s}$ and the unit vector $\mathbf{n}_i$ in the direction of the rotation axis of the torus with charge $Q_i$. We will consider the limiting case of small values for $\Delta$, $\delta_i$ and $\delta_o$. The quantity $\cos\Delta$ in (3.7) can then be approximated by $\cos\Delta \approx \cos\delta_i\cos\delta_o$. Moreover, in the calculations below $\delta_i$ and $\delta_o$ will be treated as constants. The latter assumption is only justified for small values of $\delta_i$ and $\delta_o$, too. The interesting limiting case of $\delta_i = \delta_o = 90°$ will not be treated in this work. Only the Lense-Thirring frequency $\bar{v}_{LT}^-$ for $\delta_m = 90°$ has been considered in (1.9).

Substitution of **B**(em) from (3.7) into (1.7) yields the following expression for the angular precession velocity $\mathbf{\Omega}_{io}$

$$\mathbf{\Omega}_{io} = -4\pi\beta^{-1}\frac{G^{\frac{1}{2}}Q_o}{c^2 r_o}v_o g(x)\cos\delta_i \cos\delta_o \, \mathbf{n}_o. \tag{3.8}$$

So, for $\beta = +1$ and a positive charge $Q_o$ the precession velocity $\mathbf{\Omega}_{io}$ is counter-clockwise around $\mathbf{n}_o$. In addition, the corresponding precession frequency $v_{io}$ will be given by

$$v_{io} = -Q_o' \frac{2Gm_s}{c^2 r_o} v_o g(x)\cos\delta_i \cos\delta_o, \tag{3.9}$$

where $Q_o'$ is again defined by $Q_o' \equiv (G^{\frac{1}{2}}m_s)^{-1}Q_o$.

Analogously to the calculation of $v_{io}$, the precession frequency $v_{mo}$ from the torus with total electric charge $Q_o$ acting on the mass current with total mass $m_m$ can be found. Calculation yields

$$v_{mo} = -Q_o' \frac{2Gm_s}{c^2 r_o} v_o g(x_o)\cos\delta_m \cos\delta_o, \tag{3.10}$$

where $x_o$ is defined by $x_o \equiv r_m/r_o$ and $\delta_m$ is the angle between the direction of the rotation axis of the star $\mathbf{s} = \mathbf{\Omega}_s/\Omega_s$ and the unit vector $\mathbf{n}_m$ in the direction of the rotation axis of the torus with total mass $m_m$. The function $g(x_o)$ in (3.10) has been analogously defined to $g(x)$ in (2.21) and (3.6)

$$g(x_o) \equiv \frac{2}{\pi}\left\{\frac{E(x_o)}{(1-x_o^2)}\right\}. \tag{3.11}$$

Note that the frequencies $v_{io}$ and $v_{mo}$, both originating from the charge $Q_o$, have the quantities $Q_o' \equiv (G^{\frac{1}{2}}m_s)^{-1}Q_o$, $Gm_s/(c^2 r_o)$, $v_o$ and $\cos\delta_o$ in common. For black holes, pulsars and white dwarfs the dimensionless quantity $Gm_s/(c^2 r_o)$ is smaller than unity value, so that the frequencies $v_{io}$ and $v_{mo}$ are usually smaller than $v_o$ and may therefore be denoted as low frequency QPOs.

## 4. LOW FREQUENCY QUASI-PERIODIC OSCILLATIONS, DUE TO $Q_i$

Analogously to calculation of $v_{io}$, the frequency $v_{oi}$ can be found. We now consider the circular torus containing a total charge $Q_i$, lying in another $x$-$y$ plane at distance $r_i$ from the origin $O$ (compare with figure 1). A radius vector $\mathbf{r}_o$ from $O$ to field point $F$, where a point charge $dQ_o$ is situated, is fixed by the spherical coordinates $r_o$, $\theta$ and $\varphi = 0$. The current $dQ_i/dt$ from the charge in the torus generates a vector potential in the $y$ direction at field point $F$, $A_\varphi(r_o, \theta)$, of magnitude (compare with derivation of, e.g., Jackson [14])

$$A_\varphi(r_o, \theta) = \frac{x}{c}\frac{dQ_i}{dt}\int_0^{2\pi}\frac{\cos\varphi \, d\varphi}{\left(1+x^2-2x\sin\theta\cos\varphi\right)^{\frac{1}{2}}}. \tag{4.1}$$

This integral can be evaluated in a similar way as (3.1). Analogously to the calculation of

(3.3), on finds for $A_\varphi(r_o, \theta = 90°)$

$$A_\varphi(r_o, \theta=90°) = \frac{4Q_i v_i}{c}\{K(x) - E(x)\}. \tag{4.2}$$

Combination of (3.5) and (4.2) yields for $B_\theta(r_o, \theta = 90°)$

$$B_\theta(r_o, \theta=90°) = \frac{-4Q_i v_i}{cr_o}\left\{K(x) - \frac{E(x)}{1-x^2}\right\} = \frac{2\pi Q_i v_i}{cr_o} x f(x), \tag{4.3}$$

where the function $f(x)$ has earlier been defined in (2.9). The field $\mathbf{B}$(em) from charge $Q_i$ acting on the component $\mathbf{S}_o \sin\Delta$ ($\mathbf{S}_o$ is the angular momentum of the torus with mass $m_o$ and charge $Q_o$) can be approximated by

$$\mathbf{B}(\text{em}) \approx \frac{-2\pi Q_i v_i}{cr_o} x f(x) \cos\Delta\, \mathbf{n}_i \approx \frac{-2\pi Q_i v_i}{cr_o} x f(x) \cos\delta_i \cos\delta_o\, \mathbf{n}_i, \tag{4.4}$$

where $\mathbf{n}_i$, $\Delta$, $\delta_i$ and $\delta_o$ have earlier been defined. Moreover, the approximation $\cos\Delta \approx \cos\delta_i \cos\delta_o$ has again been applied.

Substitution of $\mathbf{B}$(em) from (4.4) into (1.7) yields the following expression for the angular precession velocity $\mathbf{\Omega}_{oi}$

$$\mathbf{\Omega}_{oi} = 4\pi\beta^{-1} \frac{G^{1/2} Q_i}{c^2 r_o} v_i x f(x) \cos\delta_i \cos\delta_o\, \mathbf{n}_i. \tag{4.5}$$

Choosing $\beta = +1$ and a negative charge $Q_i$, the predicted precession velocity $\mathbf{\Omega}_{oi}$ will be counter-clockwise around $\mathbf{n}_i$. The corresponding precession frequency $v_{oi}$ will then be given by

$$v_{oi} = Q_i' \frac{2Gm_s}{c^2 r_o} v_i x f(x) \cos\delta_i \cos\delta_o, \tag{4.6}$$

where $Q_i'$ is again be defined by $Q_i' \equiv (G^{1/2} m_s)^{-1} Q_i$.

Analogously to the calculation of $v_{oi}$, the precession frequency $v_{mi}$ from the torus with total electric charge $Q_i$ acting on the mass current with total mass $m_m$ can be found. Calculation yields

$$v_{mi} = Q_i' \frac{2Gm_s}{c^2 r_m} v_i x_i f(x_i) \cos\delta_m \cos\delta_i, \tag{4.7}$$

where $x_i$ is defined by $x_i \equiv r_i/r_m$ and $\delta_m$ is the angle between the direction of the rotation axis of the star $\mathbf{s} = \mathbf{\Omega}_s/\Omega_s$ and the unit vector $\mathbf{n}_m$ in the direction of the rotation axis of the torus with total mass $m_m$. The function $f(x_i)$ has analogously been defined to $f(x)$ in (2.9) and (4.3)

$$f(x_i) = \frac{-2}{\pi x_i}\left\{K(x_i) - \frac{E(x_i)}{1-x_i^2}\right\}. \tag{4.8}$$

It is noticed that the four frequencies $v_{io}$, $v_{mo}$, $v_{oi}$ and $v_{mi}$ are all deduced from (1.7). Note that these frequencies are a unique consequence of our special interpretation of the gravitomagnetic theory.

In addition, classical *electromagnetic* precession frequencies can be calculated from (1.10). For example, the electromagnetic analogue of $v_{io}$ from (3.9), $v_{io}$(em), can be calculated by combination of (1.10) and (3.7). One obtains

$$v_{io}(\text{em}) = -Q'_i Q'_o \frac{Gm_s}{2c^2 r_o} \frac{m_s}{m_i} v_o g(x) \cos\delta_i \cos\delta_o, \qquad (4.9)$$

where $m_s$ is again the mass of the star and $m_i$ is the mass of the torus with charge $Q_i$. Analogously, from (1.10) and (4.4) an electromagnetic precession frequencies $v_{oi}$(em) can be deduced

$$v_{oi}(\text{em}) = Q'_i Q'_o \frac{Gm_s}{2c^2 r_o} \frac{m_s}{m_o} v_i x f(x) \cos\delta_i \cos\delta_o. \qquad (4.10)$$

By combining (3.9) and (4.9) the ratio between $v_{io}$ and $v_{io}$(em) can be calculated to be

$$\frac{v_{io}(\text{em})}{v_{io}} = \tfrac{1}{4} Q'_i \frac{m_s}{m_i}. \qquad (4.11)$$

Usually, $m_s$ is much larger than $m_i$ and $m_o$, respectively, so that $v_{io}$(em) and $v_{oi}$(em) are possibly out of observational range for pulsars and black holes. However, it will be shown below that $v_{io}$(em) and $v_{oi}$(em) may be observable for white dwarfs.

## 5. PARAMETER $\beta^*$

When both a magnetic induction field $\mathbf{B}_p$(gm) from gravitomagnetic origin and a field $\mathbf{B}_p$(em) from electromagnetic origin are present at the north/south pole of the pulsar, the total magnetic induction field $\mathbf{B}_p$(tot) is given by (see [8])

$$\mathbf{B}_p(\text{tot}) = \mathbf{B}_p(\text{gm}) + \mathbf{B}_p(\text{em}). \qquad (5.1)$$

According to (1.4), the direction of $\mathbf{B}_p$(gm) is antiparallel to $\mathbf{\Omega}_s$ for $\beta = +1$. It appears helpful to define the following dimensionless quantity $\beta^*$

$$\mathbf{B}_p^{\parallel}(\text{tot}) = \beta^* \mathbf{B}_p(\text{gm}). \qquad (5.2)$$

When the total field $\mathbf{B}$(tot) is only due to gravitomagnetic origin, $\mathbf{B}_p$(em) = 0 and $\beta^*$ reduces to $\beta^* = 1$. As a rule, measurements only yield $B_p$(tot), so that only an estimate for $\beta^*$ can be obtained.

As a first example, the field $\mathbf{B}$(em) at a field point $F$ generated by the total charge $Q_o$ moving in the circular torus with radius $r_o$ is calculated. The field point $F$ is now placed on the rotation axis of the torus at a position $\mathbf{r}_i$ from the origin $O$ to the point $F$ (compare with figure 1). By combining (3.4) and (3.5) one obtains

$$\mathbf{B}_{r_i}(r_i, \theta \to 0°) = \frac{2\pi Q_o v_o}{c r_o} \frac{1}{(1+x^2)^{3/2}} \mathbf{n}_o, \quad x \equiv r_i / r_o, \qquad (5.3)$$

where $\mathbf{n}_o$ is again the unit vector in the direction of the rotation axis of the torus with charge $Q_o$. In (5.3) the distance $r_i$ can be replaced by $r_s$ and from the result the following estimate for the contribution $\mathbf{B}_p^{\parallel}(Q_o)$ to the field $\mathbf{B}_p^{\parallel}(em)$ at, say, the north pole of the star can be calculated

$$\mathbf{B}_p^{\parallel}(Q_o) \approx B_{r_s}(r_s, \theta \to 0^\circ) \cos\delta_o \, \mathbf{s} = \frac{2\pi Q_o v_o}{c} \frac{r_o^2 \cos\delta_o}{\left(r_s^2 + r_o^2\right)^{3/2}} \mathbf{s}, \tag{5.4}$$

where $\delta_o$ is the angle between the direction of the rotation axis $\mathbf{s} = \mathbf{\Omega}_s/\Omega_s$ of the star and the unit vector $\mathbf{n}_o$.

Analogously, a second contribution $\mathbf{B}_p^{\parallel}(Q_i)$ to the field $\mathbf{B}_p^{\parallel}(em)$ at the north pole of the star, due to the torus with charge $Q_i$, can be calculated

$$\mathbf{B}_p^{\parallel}(Q_i) \approx B_{r_s}(r_s, \theta \to 0^\circ) \cos\delta_i \, \mathbf{s} = \frac{2\pi Q_i v_i}{c} \frac{r_i^2 \cos\delta_i}{\left(r_s^2 + r_i^2\right)^{3/2}} \mathbf{s}, \tag{5.5}$$

where $\delta_i$ is the angle between the direction of the rotation axis $\mathbf{s} = \mathbf{\Omega}_s/\Omega_s$ of the star and the unit vector $\mathbf{n}_i$.

Finally, by combining (1.3) and (1.5) a third contribution $\mathbf{B}_p^{\parallel}(Q_s)$ to the field $\mathbf{B}_p^{\parallel}(em)$ at the north pole of the star can be calculated, due to the charge $Q_s$ in the star. One obtains

$$\mathbf{B}_p^{\parallel}(Q_s) = \frac{4\pi Q_s v_s}{5 c r_s} \mathbf{s}. \tag{5.6}$$

The field of (5.6) represents the polar field of an ideal magnetic dipole located in the centre of the star, generated by a point charge $Q_s$. For a homogeneous charge distribution in the sphere it has recently been shown [9], that the polar field of (5.6) remains the same. The charge distribution in a real star, however, may be otherwise. The charge $Q_s$ may even partly reside outside the star, at the equator, for instance. Of course, these complications may affect the validity of relation (5.6).

By combining equations (1.4), (5.1), (5.4), (5.5) and (5.6) with (5.2), the following expression for $\beta^*$ can be found (a value of $\beta = +1$ has been chosen in (1.4))

$$\beta^* = 1 + \beta^*_{current} - Q_s' - \tfrac{5}{2} Q_i' \frac{v_i}{v_s} \frac{r_i^2/r_s^2 \cos\delta_i}{\left(1 + r_i^2/r_s^2\right)^{3/2}} - \tfrac{5}{2} Q_o' \frac{v_o}{v_s} \frac{r_o^2/r_s^2 \cos\delta_o}{\left(1 + r_o^2/r_s^2\right)^{3/2}}, \tag{5.7}$$

where all quantities have been defined earlier. A related expression for $\beta^*$ was previously deduced [8] for a star with a flat disk. The term $\beta^*_{current}$ in (5.7) has been added to account for a possible contribution from toroidal currents in the star (see discussion following (2.25)). For $\beta^*_{current} = -1$ the toroidal currents completely compensate the magnetic field from gravitomagnetic origin. A striking property of (5.7) is that it provides a relation between the high frequency QPOs $v_o$ and $v_i$, and the rotation frequency $v_s$ of the star.

When $v_o \gg v_s$ and $v_i \gg v_s$ the terms depending on $Q_o'$ and $Q_i'$ dominate the right hand side of $\beta^*$ in (5.7). When more high frequencies $v_i$ and $v_o$ are measured in the future, relation (5.7) may be useful to explain the magnitude of the parameter $\beta^*$. For example, more insight in the contributions to $\beta^*$ may be obtained for the binary pulsars given in table 2 of ref. [8]. Moreover, it is noticed that the expressions for $\mathbf{B}_p^{\parallel}(Q_o)$ of (5.4) and $\mathbf{B}_p^{\parallel}(Q_i)$ of (5.5) do not depend on the rotation frequency $v_s$ of the star. The observed total magnetic fields $B_p(tot)$ of the pulsars in table 2 of ref. [8] appear to be largely independent of $v_s$. Perhaps, these fields may mainly be attributed to fields like $\mathbf{B}_p^{\parallel}(Q_o)$ and/or $\mathbf{B}_p^{\parallel}(Q_i)$.

## 6. SUMMARY OF THE THEORETICAL RESULTS

In this section the main formulas obtained in this work will be summarized and a number of consequences will be discussed. Their origin has shortly been denoted, whereas all parameters have earlier been defined in this work.

### 6.1 High frequency QPOs

It will be assumed that high frequency QPO $v_i$ of (2.15) is due to an inner torus with charge $Q_i$ and radius $r_i$, whereas the high frequency QPO $v_o$ of (2.24) is due to an outer torus with charge $Q_o$ and radius $r_o$. So, $r_i < r_o$. The frequencies are

$$v_i = \frac{1}{2\pi} \left[ \frac{Gm_s}{r_i^3} \left\{ 1 - \frac{m_s}{m_i} Q_i' \left( Q_s' - x^2 f Q_o' \right) \right\} \right]^{1/2}, \quad (6.1)$$

$$v_o = \frac{1}{2\pi} \left[ \frac{Gm_s}{r_o^3} \left\{ 1 - \frac{m_s}{m_o} Q_o' \left( Q_s' + g Q_i' \right) \right\} \right]^{1/2}. \quad (6.2)$$

The expressions for the frequencies $v_i$ and $v_o$ will not quantitatively be tested in this work. When the angle $\Delta$ (defined in section 2, see also figure 2) is small, the relations (2.8) and (2.20) should be replaced by

$$Q_s \approx x^2 f(x) Q_o, \quad (6.3)$$

$$Q_s \approx -g(x) Q_i, \quad (6.4)$$

where the quantity $x$ is defined by $x \equiv r_i/r_o$. Instead of $f(x)$ in (6.3), the averaged value $\bar{f}(\bar{x}) = \frac{1}{2} \{f_0 + f(x)\} = \frac{1}{2} \{0 + f(x)\} = \frac{1}{2} f(x)$ will be used in the calculations below, whereas $g(x)$ in (6.4) will be replaced by $\bar{g}(\bar{x}) = \frac{1}{2} \{g_0 + g(x)\}$ (see section 2).

The high frequency QPO $v_m$ of (2.27) from a torus with a total mass $m_m$, lying in between the tori with charge $Q_i$ and charge $Q_o$, can be shown to be

$$v_m = \frac{1}{2\pi} \left( \frac{Gm_s}{r_m^3} \right)^{1/2} \frac{1}{1 + \frac{4\pi v_s r_s^2}{5c^2} \left( \frac{Gm_s}{r_m^3} \right)^{1/2}} = \frac{1}{2\pi} \left( \frac{Gm_s}{r_m^3} \right)^{1/2} f_s \approx \frac{1}{2\pi} \left( \frac{Gm_s}{r_K^3} \right)^{1/2} = v_K. \quad (6.5)$$

In the calculations below the relativistic factor $f_s$ usually approaches to unity value. Therefore, when frequency $v_m$ and mass $m_s$ are known, the Kepler radius $r_K$ is a satisfactory approximation for $r_m$ in most cases. It will be assumed that $r_i < r_m < r_o$.

### 6.2 Gravitomagnetic, low frequency QPOs

The gravitomagnetic precession frequencies are given by (3.9), (3.10), (4.6) and (4.7). They are all written in terms of the quantity $Gm_s/(c^2 r_o)$ or $Gm_s/(c^2 r_m)$. The concerning equations, applied to four pulsars and a black hole below, are

$$v_{mo} = -Q_o' \frac{2Gm_s}{c^2 r_o} v_o g(x_o) \cos \delta_m \cos \delta_o, \quad x_o \equiv r_m/r_o. \quad (6.6)$$

$$v_{io} = -Q'_o \frac{2Gm_s}{c^2 r_o} v_o g(x) \cos\delta_i \cos\delta_o, \quad x \equiv r_i/r_o. \tag{6.7}$$

$$v_{mi} = Q'_i \frac{2Gm_s}{c^2 r_m} v_i x_i f(x_i) \cos\delta_m \cos\delta_i, \quad x_i \equiv r_i/r_m. \tag{6.8}$$

$$v_{oi} = Q'_i \frac{2Gm_s}{c^2 r_o} v_i x f(x) \cos\delta_i \cos\delta_o, \quad x \equiv r_i/r_o. \tag{6.9}$$

It is noticed that in deriving (6.6) through (6.9) small angles $\delta_m$, $\delta_o$ and $\delta_i$ have been assumed. On the other hand no precession occurs, when $\delta_i = \delta_m = \delta_o = 0$ (see comment to (1.8) and (3.7)). If all values of $\delta$ are small, prograde motion of $Q_i$, $m_m$ and $Q_o$ around $\mathbf{s} = \mathbf{\Omega}_s/\Omega_s$ takes place. Alternatively, retrograde motion of $Q_i$, $m_m$ and $Q_o$ around $\mathbf{s}$ implies that all values of $\delta$ are about 180°.

It is noted that by combining (6.3), (6.4), (6.7) and (6.9) the following relation can be obtained, independent of both $\cos\delta_i$ and $\cos\delta_o$

$$\frac{v_{io} v_i}{v_{oi} v_o} = \frac{g(x)^2}{x^3 f(x)^2}. \tag{6.10}$$

It appears that all parameters $x$, $f(x)$ (or $\bar{f}(\bar{x})$) and $g(x)$ (or $\bar{g}(\bar{x})$) on the right hand side of (6.10) only depend on $x$. When all the frequencies on the left hand side of (6.10) are known, all parameters on the right hand side can separately be calculated by an iteration process. In the calculations below use has been made of this property of (6.7) and (6.9).

Furthermore, a remark with respect to the magnitudes of $v_{mo}$ and $v_{io}$ can be made. Assuming $r_i < r_m < r_o$, it follows from (6.6b) and (6.7b) that $x_o > x$. According to table 1, the quantity $g(x_o)$ is then larger than $g(x)$. When the angles $\delta_m$ and $\delta_o$ do not differ too much, the frequency $v_{mo}$ is larger than $v_{io}$. Finally, no sign of any of the frequencies $v_i$, $v_m$, $v_o$, $v_{mo}$, $v_{io}$, $v_{mi}$ and $v_{oi}$ is known at present. For that reason, positive signs for all frequencies will be used below.

### 6.3 Parameter $\beta^*$

According to (5.7), the following relation between the high frequency QPOs $v_o$ and $v_i$, and the rotation frequency $v_s$ of the star exists

$$\beta^* = 1 + \beta^*_{current} - Q'_s - \tfrac{5}{2} Q'_i \frac{v_i}{v_s} \frac{r_i^2/r_s^2 \cos\delta_i}{\left(1 + r_i^2/r_s^2\right)^{3/2}} - \tfrac{5}{2} Q'_o \frac{v_o}{v_s} \frac{r_o^2/r_s^2 \cos\delta_o}{\left(1 + r_o^2/r_s^2\right)^{3/2}}. \tag{6.11}$$

Note that $\delta_i = \delta_o = 0°$ in (6.11) implies prograde motion of the charges $Q_i$ and $Q_o$ around $\mathbf{s} = \mathbf{\Omega}_s/\Omega_s$, whereas $\delta_i = \delta_o = 180°$ describes retrograde motion around $\mathbf{s}$. Prograde motion of all tori will be assumed throughout this work, i.e., the angles $\delta_i$, $\delta_o$ and $\delta_m$ are small.

In favourable cases the mass $m_s$ and radius $r_s$ of the star are known, whereas the following quantities may be obtained from observations: the parameter $\beta^*$ and eight frequencies (i.e., $v_o$, $v_m$, $v_i$, $v_s$, $v_{mo}$, $v_{io}$, $v_{mi}$ and $v_{oi}$). The eight equations (6.3)–(6.9) and (6.11) then still contain ten unknown quantities (i.e., three charges: $Q_s$, $Q_o$ and $Q_i$; three radii: $r_o$, $r_m$ and $r_i$; three angles: $\delta_o$, $\delta_m$ and $\delta_i$ and the parameter $\beta^*_{current}$). Only the approximate value of $r_m$ can directly be calculated from (6.5), when the frequency $v_m$ can be assigned. Therefore, we have arbitrarily chosen one $\delta$ value, e.g., $\delta_m$ and we have taken $\delta_i = \delta_o$. Thus, choosing the value of two parameters, it appears possible to obtain the values of the remaining eight quantities. Since the quantity $\beta^*_{current}$ does not occur in the other equations, it can only calculated from (6.11).

In the calculations below all functions $f(x)$ and $g(x)$ in (6.3), (6.4) and (6.6)–(6.10) have been replaced by their averaged counterparts $\bar{f}(\bar{x}) = \frac{1}{2}\{f_0 + f(x)\} = \frac{1}{2}f(x)$ and $\bar{g}(\bar{x}) = \frac{1}{2}\{g_0 + g(x)\}$ (see section 2). In addition, the angles $\delta$ ($\delta_o$, $\delta_m$ and $\delta_i$) occurring in (6.6)–(6.11) are treated as constants. In reality, they vary as a result of the precession processes. However, in the proposed model leading to (6.6)–(6.11), it has been assumed, that the values of the angles $\delta$ are small. Since all angles $\delta$ occur as $\cos\delta$ in these equations, the introduced errors are small in that case.

## 7.  OBSERVATIONS ON PULSARS

In this section data of a number of pulsars will be compared with predictions from section 6. Only for a few pulsars seven QPO frequencies and the rotational frequency of the pulsar have been reported. These data are available for the accreting millisecond X-ray, *binary* pulsars SAX J1808.4–3658, XTE J1807–294 and IGR J00291 +5934, and for the soft gamma repeater SGR 1806–20.

A mass $m_s = 1.4\, m_\odot$ and a radius $r_s = 10^6$ cm are assumed for all discussed pulsars. The following quantities can then be calculated

$$\frac{2Gm_s}{c^2} = 0.4136\times 10^6\,\text{cm}, \quad \frac{2Gm_s}{c^2 r_s} = 0.4136. \tag{7.1}$$

Equation (7.1b) implies strong gravitational fields, so that general relativity is relevant. In the Schwarzschild space-time the radius of the relativistic innermost stable circular orbit (ISCO) is then given by $r_{ISCO} = 6Gm_s/c^2 = 1.241\times 10^6$ cm and in the extreme limit $|Q_s| = G^{1/2} m_s$ of the Reissner-Nordstrøm space-time by $r_{ISCO} = 4Gm_s/c^2 = 0.827\times 10^6$ cm.

In our model the magnitude of the radii of the tori with charge $Q_i$, electrically neutral mass $m_m$ and charge $Q_o$, $r_i$, $r_m$ and $r_o$, respectively, follow the sequence $r_i < r_m < r_o$. For all these pulsars the QPO with the highest frequency, the so-called *upper* kHz frequency, $\nu_u$ (see, e.g., ref. [1] for this notation), is identified with the frequency $\nu_i$ of the inner orbit of radius $r_i$. Analogously, the so-called *lower* kHz frequency, $\nu_l$, is identified with the frequency $\nu_m$ (approximately equal to the Kepler frequency $\nu_K$) from the torus with mass $m_m$ and radius $r_m \approx r_K$. Finally, the frequency $\nu_o$ is identified with the lowest high frequency QPO of the outer orbit of radius $r_o$.

When the radii of $r_s$ and $r_m$ do not differ too much, there is less space between the surface of the star and the torus with neutral mass $m_m$. The quality factor $Q$ (see the definition of $Q$ below) and the r.m.s. amplitude of $\nu_i$ may then be relatively high. When the radii of $r_m$ and $r_o$ do not differ much, the quality factor $Q$ and the r.m.s. amplitude of $\nu_m$ may also be high. More space may be available for the outer torus with frequency $\nu_o$. A relatively low value for the quality factor $Q$ has indeed been found for the frequency $\nu_o$. The r.m.s. amplitudes of the latter QPO are not so marginal, however.

The assignment of the remaining QPO frequencies will be given in the tables below. An estimate of the respective magnitudes of these frequencies can often be made. For example, the frequency $\nu_{mo}$ is usually larger than $\nu_{io}$, as has been discussed after (6.10). Precise values of the parameters, like $x_i$, $\bar{f}(\bar{x}_i)$, $x_o$, $\bar{g}(\bar{x}_o)$, and so on, obtained by an iteration process, have been given in the tables, since these quantities are often very sensitive to small mutual changes in their value. Of course, the real accuracy of the parameters like $x$, $r_i$, $r_o$, $Q_s'$, $Q_o'$, $Q_i'$, $\delta_i$, $\delta_m$ and $\delta_o$ is much more limited, especially the $\delta$ values.

Furthermore, calculated values for the Lense-Thirring frequency $\nu_{LT}(m_i)$ and $\nu_{LT}(m_m)$ from (1.8b) for the tori with mass $m_i$ and radius $r_i$ and with mass $m_m$ and radius $r_m$, respectively, have been added to most tables. The predicted frequencies $\nu_{LT}$ from (1.8b) or (1.9) have not yet been detected unambiguously, however. We now first consider the pulsar SAX J1808.4–3658 more in detail.

## 7.1. SAX J1808.4–3658

Data of the accreting millisecond pulsar SAX J1808.4–3658 from van Straaten *et al.* [15] are given in table 2. Since seven QPO frequencies have simultaneously been observed, including the lower kHz frequency, $v_l$, in group 3 of their tables 1–3, data of this group have exclusively been chosen. In our calculations the Lorentzian centroid frequency $v_0$ has been used, instead of the so-called characteristic frequency $v_{max} = (v_0^2 + \Delta^2)^{1/2}$ ($\Delta$ is the half-width at half maximum). Quality factors $Q \equiv v_0/(2\Delta)$ and integrated fractional r.m.s. amplitudes of the QPOs have also been given in table 2. For this pulsar the so-called hectohertz QPO, $v_{hHz}$ (see for this notation, e.g., ref. [15]), has been identified with the frequency $v_o$.

Utilizing $m_s = 1.4\ m_\odot$ and other necessary parameters, the radii $r_K$ and $r_m$ can now be calculated from (6.5). Using the frequencies $v_o$, $v_i$, equation (6.10) and other combinations of (6.6)–(6.9), accurate fits could be found between observed and calculated values of $v_{mo}$, $v_{io}$, $v_{mi}$ and $v_{oi}$ by application of an iteration process. As already has been noticed, the value for $\delta_m$ has been chosen and the approximation $\delta_i = \delta_o$ has been used. Results have been summarized in table 2 ($x_i$ and $x_o$ have been approximated by $x_i = r_i/r_K$ and $x_o = r_K/r_o$).

Calculated values for the Lense-Thirring frequencies $v_{LT}(m_i)$ and $v_{LT}(m_m)$ from (1.8b), have also been added to table 2. Note that frequencies $v_{LT}(m_i)$ and $v_{oi}$ coincide, when $r_s = 1.3 \times 10^6$ cm is substituted into (1.8b).

Table 2. Frequencies, quality factors $Q$ and integrated fractional r.m.s. amplitudes of the pulsar SAX J1808.4–3658 are summarized. Relative radii ($x$, $x_o$ and $x_i$), radii ($r_i$, $r_K$, $r_m$ and $r_o$), relative charges $Q_s'$, $Q_o'$ and $-Q_i'$ ($Q'$ is defined by $Q' \equiv (G^{1/2}m_s)^{-1}Q$), factors $f(\bar{x})$, $f(\bar{x_i})$, $\bar{g}(\bar{x})$ and $\bar{g}(\bar{x_o})$, and angles $\delta_i$ and $\delta_o$ are calculated. See text for comments.

| $v_{max}$[a] (Hz) | $v_0$[b] (Hz) | $Q$[c] | r.m.s.[d] (%) | $x$ | $R \times 10^6$ (cm) | $Q'$ | $f(\bar{x})$[e] | $\bar{g}(\bar{x})$[e] | $\delta$ (°) |
|---|---|---|---|---|---|---|---|---|---|
| $v_u$ 685.1 | $v_i$ 682.4 | 5.60 | 8.78 | | $r_i$ 2.220 | $-Q_i'$ 0.2371 | | | $\delta_i$ 26.24 |
| $v_l$ 503.6 | $v_m$ 503.3 | 14.26 fixed | 2.94 | | $r_m$ 2.643 $r_K$ 2.649 | $Q'$ 0 | | | $\delta_m$ 5 |
| $v_s$ 401 | $v_s$ 401 | large | | | $r_s$ 1 | $Q_s'$ 0.3328 | | | |
| $v_{hHz}$ 329 | $v_o$ 189 | 0.35 | 11.66 | | $r_o$ 3.097 | $Q_o'$ 1.623 | | | $\delta_o$ 26.24 |
| $v_h$ 74.9 | $v_{mo}$ 73.6 | 2.66 | 6.79 | $x_o$ 0.85529 | $r_o$ 3.097 | $Q_o'$ 1.623 | | $\bar{g}(\bar{x_o})$ 2.0101 | $\delta_m=5$ $\delta_o$ |
| $v_{LF}$ 47.28 | $v_{io}$ 46.27 | 2.38 | 7.17 | $x$ 0.71690 | $r_o$ 3.097 | $Q_o'$ 1.623 | | $\bar{g}(\bar{x})$ 1.4035 | $\delta_i=\delta_o$ 26.24 |
| $v_b$ 16.02 | $v_{mi}$ 15.02 | 1.35 | 8.7 | $x_i$ 0.83820 | $r_m$ 2.643 | $-Q_i'$ 0.2371 | $f(\bar{x_i})$ 0.7938 | | $\delta_m=5$ $\delta_i$ |
| $v_{b2}$ 10.4 | $v_{oi}$ 4.97 | 0.272 | 12.55 | $x$ 0.71690 | $r_o$ 3.097 | $-Q_i'$ 0.2371 | $f(\bar{x})$ 0.3988 | | $\delta_i=\delta_o$ 26.24 |
| | $v_{LT}(m_i)$ 3.03 | | | | $R=r_i$ 2.220 | | | | |
| | $v_{LT}(m_m)$ 1.80 | | | | $R=r_m$ 2.643 | | | | |

[a] Characteristic frequencies $v_{max}$, taken from [15]. [b] The centroid frequencies $v_0$ have been calculated from data given in [15]. [c] Quality factors $Q$, taken from [15]. [d] Ref. [15]. [e] Definitions of these quantities have been given in section 2.

As an illustration, the precessing tori around the star with charge $Q_i$, mass $m_m$ and charge $Q_o$, respectively, are schematically given in figure 3.

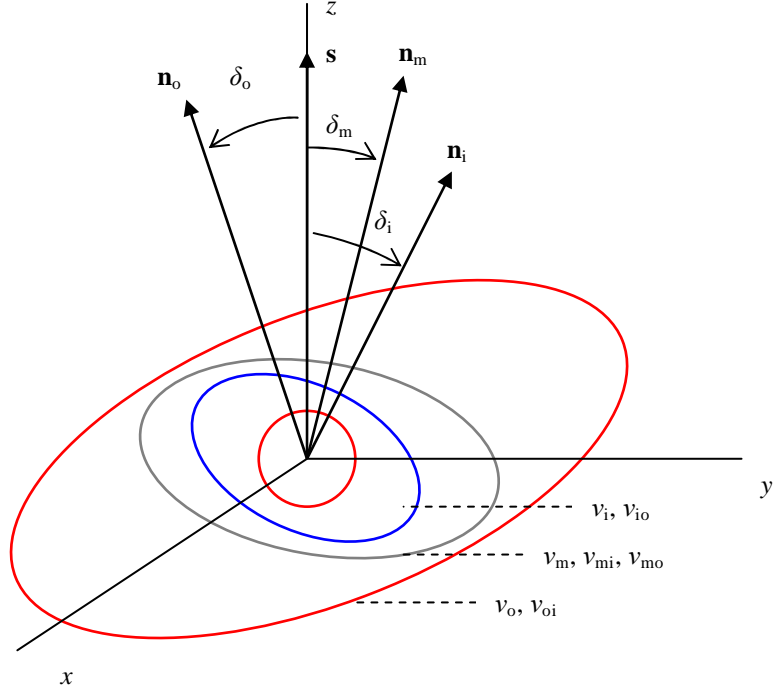

Figure 3. Rotation axes of a pulsar like SAX J1808.4–3658. The unit vector of the rotation axis of the star is $\mathbf{s} \equiv \mathbf{\Omega}_s/\Omega_s$. The unit vectors in the direction of the rotation axes of the circular tori with charge $Q_i$ (blue), mass $m_m$ (grey) and charge $Q_o$ (red) are given by $\mathbf{n}_i$, $\mathbf{n}_m$ and $\mathbf{n}_o$, respectively. The angle between $\mathbf{s}$ and $\mathbf{n}_i$ is given by $\delta_i$, and so on (not drawn to scale). The frequencies of the various tori have also been denoted.

No value for the parameter $\beta^*$ for SAX J1808.4–3658 is yet available. Therefore, in order to be able to calculate $\beta^*_{\text{current}}$ from (6.11), an estimate for $\beta^*$ must be made. Only for the isolated, millisecond pulsar B1821–24 ($\nu_s = 328$ Hz) a value $\beta^* = 2\times 10^{-5}$ has been extracted from *electron* cyclotron resonance spectral features (see ref. [8] for a discussion of this result). For that reason, we will use the value $\beta^* \approx 0$ for SAX J1808.4–3658 and the other millisecond pulsars. According to (5.2), this assumption implies that the component of magnetic field at the pole $B_p^{\parallel}(\text{tot})$ is much smaller than the gravitomagnetic field $B_p(\text{gm})$. For SAX J1808.4–3658 an absolute value for $B_p(\text{gm}) = 2.4\times 10^{16}$ G can be calculated from (1.4), using the values $m_s = 1.4\, m_\odot$, $r_s = 10^6$ cm, $\nu_s = 401$ Hz and $\beta = +1$. This value of $B_p(\text{gm})$ can be compared with the magnetic field at the north pole of the pulsar, $B_p(\text{sd})$, deduced from the standard magnetic dipole radiation model. The latter quantity can be calculated from $B_p(\text{sd}) = 3.2\times 10^{19}(-\dot{\nu}_s/\nu_s^3)^{\frac{1}{2}}$ (see, e.g, ref. [8] for this expression and a discussion of both $B_p(\text{gm})$ and $B_p(\text{sd})$). Substitution of the values $\nu_s = 401$ Hz and $\dot{\nu}_s = -2\times 10^{-13}$ Hz.s$^{-1}$ from Chakrabarty *et al.* [16] yields a value $B_p(\text{sd}) = 1.8\times 10^9$ G. The latter value also suggests that $\beta^* \approx 0$. It is noticed, that a detailed discussion of the value of the parameter $\beta^*$ for different kinds of pulsars, like binary millisecond pulsars, isolated pulsars and so on, has been given in ref. [8].

Substitution of $\beta^* = 0$ and the other necessary data from table 2 into equation (6.11) leads to the following value for $\beta^*_{\text{current}}$

$$\beta^*_{\text{current}} = -1 + 0.333(\text{from}\, Q_s) - 0.309(\text{from}\, Q_i) + 0.477(\text{from}\, Q_o) = -0.499. \quad (7.2)$$

In view of the uncertain value of $\beta^*$, the value of $\beta^*_{\text{current}}$ can only be regarded as an estimate. The value $\beta^*_{\text{current}} = -0.499$ can be compared with the asymptotic value $\beta^*_{\text{current}} = -1$ (see section 5 and discussion of a number of binary millisecond pulsars in ref. [8]).

Note that the product of $Q' \equiv (G^{1/2} m_s)^{-1} Q$ and $m_s$ in the gravitomagnetic precession frequencies (6.6)–(6.9) is independent of $m_s$. Thus, when the mass of SAX J1808.4–3658 would be $m_s = 1.4/0.875\, m_\odot = 1.6\, m_\odot$ instead of $1.4\, m_\odot$, the charge $Q_o' = 1.623$ would reduce to $0.875 \times 1.623 = 1.420$, the charge $Q_i' = -0.2371$ to $-0.2074$ and the charge $Q_s' = 0.3328$ to $0.2912$, respectively.

Apart from the QPO frequencies in table 2, the origin of other features may become more clear by the proposed model. For example, some aspects of the behaviour of a thermonuclear X-ray burst of SAX J1808.4–3658 on 18 October 2002, discussed by Chakrabarty *et al.* [16], may be explained. These authors adopt a decoupled surface burning layer that expands and contracts during a burst on a time scale of about 40 seconds. From observations they found a steep rise of the burst frequency from about 395 Hz to about 402 Hz, followed by a slow decrease to about 395 Hz (see their figure 1).

Later on, Poutanen and Gierliński [17] gave an analysis of the observed X-ray pulse profiles. They assumed that the X-ray emitting spots are located at the magnetic poles of the pulsar, at a small distance from the spin axes. As a consequence, a nearly sinusoidal modulation of the X-rays is generated. They also deduced an angle of about 11° between the spin axis and the unit vector pointing from the centre of the pulsar to the emitting spot. This result is fair agreement with our assumption that the directions of the spin axis $\mathbf{\Omega}_s$ and $\mathbf{M}(\text{gm})$ from (1.1) are antiparallel for $\beta = +1$, whereas $\mathbf{\Omega}_s$ and $\mathbf{M}(\text{em})$ from (1.5) are antiparallel for a positive charge $Q_s$.

We now first consider the expansion of the burning layer more in detail. If the parameter $x \equiv r_i/r_o$ increases during this stage ($r_i$ increases more than $r_o$), the equilibrium between the charges $Q_s$, $Q_i$ and $Q_o$ will be distorted. In order to restore charge equilibrium, some negative charge may flow from the pole of the star and/or some positive charge to the pole (see comment to (2.25)). The quantity of positive charge rapidly moves through the magnetic induction field $\mathbf{B}(\text{tot})$ *near* the magnetic pole of the star. When the direction of component of $\mathbf{B}(\text{tot})$ parallel to the surface of the star points to the pole, the Lorentz force may accelerate the positive charge entering the star to a frequency somewhat higher than the rotational frequency of the star, $\nu_s = 401$ Hz. Thus, during the expansion of the burning layer the Lorentz force might enhance the frequency of the positive charge slightly outside the star from 395 Hz to about 402 Hz, whereas during contraction of the burning layer some positive charge may return from the star. The Lorentz force may then reduce the frequency of the positive charge to its pre-burst value of 395 Hz (It is noticed that electrons and protons, for example, will flow in opposite directions, but protons may affect the rotational frequencies more strongly).

Likewise, during expansion of the burning layer positive charge might rapidly move through the magnetic induction field $\mathbf{B}_{\text{eq}}^{\|}(\text{tot})$ at the equator of the star. In order to calculate this field $\mathbf{B}_{\text{eq}}^{\|}(\text{tot})$ the same line of reasoning is followed as in the derivation of (5.7). Analogous to the derivation of (5.6), the field at the equator from $Q_s$ has been deduced from an *ideal* magnetic dipole model. The following expression for this field $\mathbf{B}_{\text{eq}}^{\|}(\text{tot})$ at distance $r_s$ from the centre of the star can be found

$$\mathbf{B}_{\text{eq}}^{\|}(\text{tot}) = \left\{ 1 + \beta^*_{\text{current}} - Q_s' + \tfrac{5}{2} Q_i' \frac{\nu_i}{\nu_s} \frac{r_i^2 (2r_i/r_s - 1)\cos\delta_i}{r_s^2 (1 + r_i/r_s)^4} \right. \\ \left. + \tfrac{5}{2} Q_o' \frac{\nu_o}{\nu_s} \frac{r_o^2 (2r_o/r_s - 1)\cos\delta_o}{r_s^2 (1 + r_o/r_s)^4} \right\} \frac{2\pi G^{1/2} m_s \nu_s}{5 c r_s} \mathbf{s}, \tag{7.3}$$

where all parameters have been defined before. Introduction of all necessary quantities from table 2 and (7.2) and insertion of (1.4) yields

$$\mathbf{B}_{eq}^{\parallel}(\text{tot}) = 0.329 \frac{2\pi G^{\frac{1}{2}} m_s \nu_s}{5 c r_s} \mathbf{s} = -\tfrac{1}{2} 0.329 \, \mathbf{B}_p(\text{gm}). \qquad (7.4)$$

It has recently been shown [9], however, that the equatorial field of a sphere with homogeneous mass or charge density may be neglected. Then, one might assume $(1 + \beta^*_{current} - Q_s') \approx 0$ in relation (7.3), so that $\mathbf{B}_{eq}^{\parallel}(\text{tot})$ changes into $\mathbf{B}_{eq}^{\parallel}(\text{tot}) \approx -\tfrac{1}{2} 0.161 \, \mathbf{B}_p(\text{gm})$.

As an example, the existence and sign of $\beta^*_{current}$ in (5.7) may be traced to previous inflow of positive charge into the star at the equator, leading to the total positive charge $Q_s$ of the star. The Lorentz force may have deflected the positive charge entering the superconducting neutron star (see for the direction and the strength of the magnetic induction field $\mathbf{B}_{eq}^{\parallel}(\text{tot})$ of (7.4)), so that a toroidal current has been generated in the star. Therefore, an additional contribution $\beta^*_{current}$ with negative sign has to be added to $\beta^*$.

Another example of an unexplained feature in the observations of J1808.4–3658 has been detected by Wijnands *et al.* [18]. They found an additional narrow QPO frequency near 410 Hz. In addition, Van Straaten *et al.* [15] found this 410 Hz QPO in group 5, 6, 9, 10 and 11 (see their table 5). The explanation of this QPO may be as follows. The stages of group 5, 6, and so on, take place after the nearest approach of the three tori during the stage of group 3 ($\nu_o$ is at maximum and $\nu_i$ just passed its maximum). As a consequence of liberated heath, e.g., from nuclear and electric origin, expansion of the tori may occur. Instead of 40 seconds, this process takes place on a time scale of days. Again the parameter $x$ may increase and the equilibrium between the charges $Q_s$, $Q_i$ and $Q_o$ will be distorted. In order to restore the charge equilibrium, some positive charge may flow to the star at the equator (see again comment to (2.25)). In this case the Lorentz force may also enhance the frequency of the positive charge slightly outside the star to about 410 Hz, higher than the rotational frequency of the star, $\nu_s = 401$ Hz.

**7.2. XTE J1807–294**

As a second example, data obtained by Zhang *et al.* [19] for the accreting millisecond pulsar XTE J1807–294 are summarized in table 3. Data of group 5 of their table 2 from the outburst decay in March 2003 have been chosen for two reasons. First, seven QPO frequencies have simultaneously been observed for this group. Secondly, this group of frequencies contains the highest value for $\nu_m$. Then, the torus with mass $m_m$ possesses the smallest value for $r_m$ (see (6.5)). Comparable results for XTE J1807–294 were reported by Linares *et al.* [20] in group H of their table 3 from observations between February and July 2003. However, they did not report the two lowest QPO frequencies, necessary for our analysis. This difference may demonstrate the intermittent character and weakness of some QPO signals. For pulsar XTE J1807–294 the same procedure has been applied as in the previous case.

Calculated values for the Lense-Thirring frequencies $\nu_{LT}(m_i)$ and $\nu_{LT}(m_m)$ from (1.8b), have also been added to table 3. Although very weak QPO signals might be present in the reported spectra in figure 4 of ref. [19] and in figure 6H of ref. [20], no unambiguous assignments for these frequencies has been made up to now.

Choosing again $\beta^* = 0$ and introducing the other necessary data from table 3, the following value for $\beta^*_{current}$ can be calculated from (6.11)

$$\beta^*_{current} = -1 + 0.261(\text{from } Q_s) - 0.222(\text{from } Q_i) + 0.224(\text{from } Q_o) = -0.737. \qquad (7.5)$$

It appears that the value for $\beta^*_{current}$ of (7.5) for XTE J1807–294 is closer to the asymptotic value – 1 for millisecond pulsars than the value for $\beta^*_{current}$ of (7.2) (see again ref. [8]).

It is noticed that in the derivations of (3.9) and (4.6), leading to $v_{io}$ and $v_{oi}$, respectively, small values for $\delta_i$ and $\delta_o$ have been adopted. Table 3 shows, however, that the found values for $\delta_i$ and $\delta_o$ are not small. In addition, line charges $Q_i$ and $Q_o$ in two tori have been assumed. However, when the charges $Q_i$ and $Q_o$ reside in an inner and outer belt, respectively, the system may also be stable. The belts can be compared with the observed inner and outer Van Allen radiation belts around the Earth. Thus, the results of table 3 may remain approximately valid.

Table 3. Data and calculated parameters for XTE J1807–294. See caption of table 2 and text for comments.

| $v_{max}$ [a] (Hz) | $v_0$ [b] (Hz) | $Q$ [c] | r.m.s. [d] (%) | $x$ | $R \times 10^6$ (cm) | $Q'$ | $\bar{f}(\bar{x})$ [e] | $\bar{g}(\bar{x})$ [e] | $\delta$ (°) |
|---|---|---|---|---|---|---|---|---|---|
| $v_u$ 545.3 | $v_i$ 544.9 | 13.5 | 10.4 | | $r_i$ 2.689 | $-Q_i'$ 0.1665 | | | $\delta_i$ 52.28 |
| $v_l$ 360 | $v_m$ 358 | 4.6 | 9.1 | | $r_m$ 3.321 $r_K$ 3.324 | $Q'$ 0 | | | $\delta_m$ 5 |
| $v_s$ 191 | $v_s$ 191 | large | | | $r_s$ 1 | $Q_s'$ 0.2611 | | | |
| $v_{hHz}$ 165 | $v_o$ 134 | 0.69 | 9.8 | | $r_o$ 3.472 | $Q_o'$ 0.8164 | | | $\delta_o$ 52.28 |
| $v_h$ 39.2 | $v_{mo}$ 38.3 | 2.24 | 9.0 | $x_o$ 0.95748 | $r_o$ 3.472 | $Q_o'$ 0.8164 | | $\bar{g}(\bar{x_o})$ 4.8216 | $\delta_i$ 52.28 $\delta_m$=5 |
| $v_b$ 11.9 | $v_{io}$ 7.65 | 0.42 | 11.4 | $x$ 0.77464 | $r_o$ 3.472 | $Q_o'$ 0.8164 | | $\bar{g}(\bar{x})$ 1.5682 | $\delta_i = \delta_o$ 52.28 |
| $v_{LFN}$ ($v_{b2}$) 3.66 | $v_{mi}$ 3.63 | 3.6 | 2.5 | $x_i$ 0.80905 | $r_m$ 3.321 | $-Q_i'$ 0.1665 | $\bar{f}(\bar{x_i})$ 0.6522 | | $\delta_i$ 52.28 $\delta_m$=5 |
| $v_{LFN/2}$ 1.68 | $v_{oi}$ 1.67 | 4.7 | 2.0 | $x$ 0.77464 | $r_o$ 3.472 | $-Q_i'$ 0.1665 | $\bar{f}(\bar{x})$ 0.5329 | | $\delta_i = \delta_o$ 52.28 |
| | $v_{LT}(m_i)$ 0.81 | | | | $R = r_i$ 2.689 | | | | |
| | $v_{LT}(m_m)$ 0.43 | | | | $R = r_m$ 3.321 | | | | |

[a] Characteristic frequencies $v_{max}$, taken from [19]. [b] The centroid frequencies $v_0$ have been calculated from data given in [19]. [c] Quality factors $Q$, taken from [19]. [d] Ref. [19]. [e] Definitions of these quantities have been given in section 2.

It is noticed, that data in table 3 are also compatible with an alternative set of parameters $x$, $r$, $\delta$, $\bar{f}(\bar{x})$, $\bar{g}(\bar{x})$ and charges $Q$, when the frequencies 544.9 Hz and 134 Hz are attributed to $v_o$ and $v_i$, respectively. In that case, a value $x = 0.9955$ is calculated. For the data of SAX J1808.4–3658 such an alternative set is also possible with $v_o = 682.4$ Hz, $v_i = 189$ Hz and $x = 0.9465$. The radii of $r_i$, $r_m$ and $r_o$ are then lying close together. Although these alternatives are formally possible, they seem less probable.

In addition, a remark with respect to the inner radius $r_i$ can be made. The radius $r_i = 2.69 \times 10^6$ cm of pulsar XTE J1807–294 (see table 3) is somewhat larger than the corresponding Kepler radius of $2.51 \times 10^6$ cm. Comparison with the pulsar SAX J1808.4–3658 shows, that the radius $r_i = 2.22 \times 10^6$ cm of the latter pulsar (see table 2) is also slightly larger than the corresponding Kepler radius of $2.16 \times 10^6$ cm. These results suggest that the charge dependent contribution to the right hand side of (2.15) is relatively unimportant for both stars (see also comment to (2.15)).

### 7.3. IGR J00291+5934

As a third example, data reported by Linares *et al.* [21] for the accreting millisecond pulsar (or black hole?) IGR J00291+5934 are summarized in table 4. During the 2004 outburst two sets (A1 and A2 in their table 3) of seven QPO frequencies have simultaneously been observed, although a fit with six frequencies was also possible. This difference may again illustrate that a weak QPO signal can easily be missed. Data of set A1 have been chosen, since this series contains the highest frequencies, including the highest value for $v_m$. Then, the torus with mass $m_m$ possesses the smallest value for $r_m$ (see (6.5)). In order to obtain the results of table 4, a procedure has been applied similar to the previous cases.

Note that the high QPO frequencies $v_i$ and $v_o$ of IGR J00291+5934 are an order of magnitude smaller than their counterparts of SAX J1808.4–3658 and XTE J1807–294. According to (6.6)–(6.9), the predicted low frequency QPOs $v_{mi}$, $v_{oi}$, $v_{mo}$ and $v_{io}$ are then relatively small. This prediction is in agreement with the found frequencies of table 4. In addition, the rotational frequency $v_s$ is much larger than $v_i$ and $v_o$ in this case. Furthermore, $\Delta v = v_i - v_m = 61.0$ Hz, so that $\Delta v/v_s = 0.10$, far less than unity value. Therefore, beat-frequency models are not confirmed by these data (see, e.g., review of van der Klis [1, § 2.8.4]).

Choosing again $\beta^* = 0$ and introducing the other necessary data from table 4, the following value for $\beta^*_{current}$ can be calculated from (6.11)

$$\beta^*_{current} = -1 + 0.439(\text{from } Q_s) - 0.002(\text{from } Q_i) + 0(\text{from } Q_o) = -0.563. \quad (7.6)$$

Note that the terms depending on the charges $Q_i$ and $Q_o$ in the right hand side of (7.6) can now be neglected. Otherwise stated, the contributions from charges $Q_i$ and $Q_o$ to $\beta^*$ of (6.11) are almost negligible. The value $\beta^*_{\text{current}} = -0.563$ can again be compared with the asymptotic value $\beta^*_{\text{current}} = -1$, discussed in ref. [8] for a number of millisecond pulsars.

Table 4. Data and calculated parameters for IGR J00291+5934. See caption of table 2 and text for comments.

| $v_{max}$ [a] (Hz) | $v_0$ [b] (Hz) | $Q$ [c] | r.m.s. [d] (%) | $x$ | $R \times 10^6$ (cm) | $Q'$ | $f(\bar{x})$ [e] | $\bar{g}(\bar{x})$ [e] | $\delta$ (°) |
|---|---|---|---|---|---|---|---|---|---|
| $v_s$ 598.88 | $v_s$ 598.88 | large | | | $r_s$ 1 | $Q_s'$ 0.4386 | | | |
| $v_u$ 66.3 | $v_i$ 66.3 | fixed | 15.6 | | $r_i$ 43.79 | $-Q_i'$ 0.2992 | | | $\delta_i$ 15.57 |
| $v_{low}$ 5.3 | $v_m$ 5.3 | fixed | 22.3 | | $r_m = r_K$ 55.13 | $Q'$ 0 | | | $\delta_m$ 40 |
| $v_h$ 0.71 | $v_o$ 0.71 | fixed | 24.0 | | $r_o$ 59.03 | $Q_o'$ 1.773 | | | $\delta_o$ 15.57 |
| $v_{2b}$ 0.078 | $v_{mi}$ 0.052 | 0.44 | 17.8 | $x_i$ 0.79432 | $r_m = r_K$ 55.13 | $-Q_i'$ 0.2992 | $f(\bar{x_i})$ 0.5961 | | $\delta_m = 40$ $\delta_i$ |
| $v_{QPO}$ 0.0436 | $v_{oi}$ 0.0430 | 2.9 | 8.3 | $x$ 0.74189 | $r_o$ 59.03 | $-Q_i'$ 0.2992 | $f(\bar{x})$ 0.4494 | | $\delta_i = \delta_o$ 15.57 |
| $v_{QPO/2}$ 0.0225 | $v_{mo}$ 0.0223 | 3.6 | 6.5 | $x_o$ 0.93400 | $r_o$ 59.03 | $Q_o'$ 1.773 | | $\bar{g}(\bar{x_o})$ 3.4256 | $\delta_m = 40$ $\delta_o$ |
| $v_{1b}$ 0.016 | $v_{io}$ 0.012 | 0.53 | 12.8 | $x$ 0.74189 | $r_o$ 59.03 | $Q_o'$ 1.773 | | $\bar{g}(\bar{x})$ 1.4659 | $\delta_i = \delta_o$ 15.57 |
| | $v_{LT}(m_i)$ (mHz) 0.59 | | | | $R = r_i$ 43.79 | | | | |

[a] Characteristic frequencies $v_{max}$, taken from [21]. [b] The centroid frequencies $v_0$ have been calculated from data given in [21]. [c] Quality factors $Q$, taken from [21]. [d] Ref. [21]. [e] Definitions of these quantities have been given in section 2.

It is noticed that the opposite sign of *all* charges of $Q_s$, $Q_i$ and $Q_o$ given in tables 2, 3 and 4, respectively, is compatible with the same set of parameters $x$, $r$, $\delta$, $f(\bar{x})$ and $\bar{g}(\bar{x})$ in those tables. The choice of the opposite the sign of all charges, however, leads to different values for $\beta^*_{\text{current}}$: $-1.50$ for SAX J1808.4–3658, $-1.26$ for XTE J1807–294 and $-1.44$ for IGR J00291+5934, respectively. Furthermore, it is noted that the choice $v_i = 0.71$ Hz and $v_o = 66.3$ Hz for IGR J00291+5934 is incompatible with relation (6.10). Substitution of the frequencies shows that the left hand side of (6.10) is smaller than unity value, whereas the right hand side predicts a result larger than unity.

In addition, it is noticed that the velocity $v_i = 2\pi v_i r_i$ (see (2.13)) is relativistic: $v_i/c = 2\pi \times 66.3 \times 4.379 \times 10^7 / 2.9979 \times 10^{10} = 0.609$. In that case the derivation of (2.13) may be modified (see comment following the derivation of (2.13)). Since the frequency $v_i$ is introduced as an empirical parameter into the calculations of table 4, the results in this table are not affected by this complication.

Furthermore, it is noticed that the found radii $r_i$, $r_m$ and $r_o$ are an order of magnitude larger than the corresponding radii for the pulsars in tables 2, 3 (and 5). Since the differences between the radii are larger in the present case, the charges $Q_i$ and $Q_o$, and the mass $m_m$ will probably possess a larger spread $\Delta r_i$, $\Delta r_o$ and $\Delta r_m$, respectively, than the other discussed pulsars. As a result, the quality factors of the high frequencies $v_i$, $v_m$ and $v_o$ may be reduced. However, the r.m.s amplitudes may remain largely unaffected. Comparison of the data in table 4 with the corresponding data in tables 2, 3 (and 5) seems to confirm this explanation.

Alternatively, the relativistic precession model proposed by Stella and Vietri [3], and Stella, Vietri and Morsink [4] may be used in order to explain the data. According to this model, the motion of neutral mass in nearly circular orbits is described in terms of the *upper* frequency QPO $v_u$ and the *lower* frequency QPO $v_l$, or the so-called periastron precession frequency $v_{\text{per}}$ ($v_l = v_{\text{per}}$). The frequency $v_{\text{per}}$ is defined by $v_{\text{per}} \equiv v_u - v_r$, where $v_r$ is the so-called epicyclic frequency. When contributions from the angular momentum $S$ of the star are neglected (compare with (2.27), $v_u$ and $v_r$ are, respectively, given by

$$v_u \approx \frac{1}{2\pi}\left(\frac{Gm_s}{r_u^3}\right)^{1/2}, \quad v_r \approx v_u\left(1 - \frac{6Gm_s}{c^2 r_u}\right)^{1/2}. \tag{7.7}$$

For a nearly circular orbit the coordinate distance may be approximated by radius $r_u$. Using (7.7) and identification of $v_u$ with 66.3 Hz and $v_l$ with 5.3 Hz (see table 4) then leads to $r_u = 11.5 \times 10^6$ cm and $m_s = 1.99\ m_\odot$ (in our calculations we assumed $m_s = 1.4\ m_\odot$). In the model of refs. [3, 4] a third lower frequency is attributed to the Lense-Thirring precession. The latter frequency may be calculated from (1.9). In table 4 the Lense-Thirring frequencies $v_{\text{LT}}(m_i)$ from (1.8b) has been added. At present, no definite identifications can be made, however. Moreover, it has been suggested by Linares *et al.* [21] that IGR J00291+5934 might even be a black hole.

### 7.4. SGR 1806–20

Further, ten different QPO frequencies for the soft gamma repeater SGR 1806–20, discovered by Strohmayer and Watts [22] are given in table 5, together with additional data. The frequencies were extracted from measurements of the 27[th] December 2004 giant flare. The two QPO frequencies near 626 Hz and the three frequencies near 92.7 Hz may be attributed to $v_m$ and $v_{mi}$, respectively. As an example, the two values for $v_m$, 626.46 Hz and 625.5 Hz, may then be caused by slightly different values of $r_m$ (see (6.5)). In our calculations we have chosen the values $v_m = 625.5$ Hz and $v_{mi} = 92.7$ Hz. The total number of different basic QPO frequencies is then reduced to seven. Note that the rotational frequency $v_s$ of SGR 1806–20 is much smaller than $v_i$ and $v_o$. In order to obtain the results of table 5, a similar procedure has been applied as above.

For SGR 1806–20 a value for $β^*$ can be deduced from observations. From *proton* cyclotron resonance spectral features an (absolute) value $B_p(tot) = 1.0 \times 10^{15}$ G for the observed magnetic field at the pole of SGR 1806–20 has been calculated. However, if the cyclotron features are attributed to *electron* cyclotron resonance a value of $B_p(tot) = 5.6 \times 10^{11}$ G is obtained. For the gravitomagnetic field $B_p(gm)$ an (absolute) value of $8.0 \times 10^{12}$ G can be calculated from (1.4), using or the values $m_s = 1.4\ m_\odot$, $r_s = 10^6$ cm, $v_s = 0.133$ Hz and $β = +1$. Application of the first interpretation yields a value of $β^* = \pm 125$ from (5.2). In addition, a magnetic field $B_p(sd) = 3.2 \times 10^{19} (-\dot{v}_s/v_s^3)^{1/2}$ can be deduced from the standard magnetic dipole radiation model. Substitution of the values $v_s = 0.133$ Hz and $\dot{v}_s = -8.69 \times 10^{-12}$ Hz.s$^{-1}$ from Woods *et al.* [23] yields a value $B_p(sd) = 1.9 \times 10^{15}$ G. A discussion of the values for $B_p(tot)$, $B_p(gm)$ and $B_p(sd)$ has earlier been given in ref. [8].

Substitution of $β^* = \pm 125$ and other necessary data from table 5 in (6.11), then yields the following values for $β^*_{current}$

$$β^*_{current} = -1 - 0(Q_s) + 1483(Q_i) - 332(Q_o) \pm 125(β^*) = +1275 \text{ or } +1025. \qquad (7.8)$$

Note that the term with charge $Q_i$ dominates the magnitude of the parameter $β^*$ in this case. Possibly, the sign and existence of $β^*_{current}$ in (6.11) may be traced to previous outflow of positive charge from the star at the equator, leading to a negative charge $Q_s$. The Lorentz force (in this case follows $\mathbf{B}_{eq}^{\parallel}(tot) = -794\ \mathbf{B}_p(gm)$ from (7.3)), may have deflected the positive charge leaving the superconducting neutron star, so that a large toroidal current $β^*_{current}$ (with positive sign) will be generated in the star.

In addition, it is again noticed that the opposite sign of *all* charges of $Q_s$, $Q_i$ and $Q_o$ given in table 5 is also compatible with the same set of parameters $x$, $r$, $δ$, $f(\bar{x})$ and $\bar{g}(\bar{x})$ in that table. The choice of the opposite the sign of all charges, however, leads to a different value for $β^*_{current}$: $-1277$ or $-1027$. Furthermore, it is noted that the choice $v_i = 150.3$ Hz and $v_o = 1837$ Hz for SGR 1806–20 is incompatible with relation (6.10). Substitution of the frequencies shows that the left hand side of (6.10) is smaller than unity value, whereas the right hand side predicts a result larger than unity.

Some remarks with respect to the inner radius $r_i$ are in order. For both SAX J1808.4–3658 and XTE J1807–294 the radii $r_i$ do not much differ from the corresponding Kepler radii, but for SGR 1806–20 the inner radius $r_i = 2.15 \times 10^6$ cm (see table 5) is substantially larger than the corresponding Kepler radius of $1.12 \times 10^6$ cm. In the relativistic precession model of Stella, Vietri and Morsink [4] the *highest* QPO frequency is identified with the frequency $v_m$ of (2.27) (for SGR 1806–20 the angular momentum contribution can be neglected, so that $v_m \approx v_K$). In that case, the Kepler value of $1.12 \times 10^6$ cm is smaller than the radius of the relativistic innermost stable circular orbit in the Schwarzschild space-time, $r_{ISCO} = 6Gm_s/c^2 = 1.24 \times 10^6$ cm. Summing up, our identification of the *second highest* QPO frequency 625.5 Hz with the frequency $v_m$ seems to be reasonable.

An alternative model for the QPOs for SGR 1806–20 in terms of global torsional frequencies has been discussed by Strohmayer and Watts [22]. The latter model has, however, been criticized by Levin [24].

Another aspect of a charged star deserves special attention. The potential energy $U$ of a star with charge $Q_s$ generating an electric field $E$ at distance $R$ is approximately given by

$$U_E = \frac{4\pi}{3} R^3 \frac{E^2}{4\pi} = \tfrac{1}{3} R^3 \frac{Q_s^2}{R^4} = \tfrac{1}{3} \frac{(Q_s')^2 G m_s^2}{R}, \qquad (7.9)$$

where all parameters have been defined before. For a mass $m_s = 1.4\ m_\odot$, a radius $R = r_s =$

Table 5. Data and calculated parameters for SGR 1806–20. See caption of table 2 and text for comments.

| $\nu_0$[a] (Hz) | $\Delta$[b] (Hz) | $Q$[c] | r.m.s.[d] (%) | $x$ | $R \times 10^6$ (cm) | $Q'$ | $f(\bar{x})$[e] | $\bar{g}(\bar{x})$[e] | $\delta$ (°) |
|---|---|---|---|---|---|---|---|---|---|
| $\nu_i$ 1837 | 4.7 | 195 | 18.0 | | $r_i$ 2.154 | $Q_i'$ 0.1291 | | | $\delta_i$ 16.16 |
| $\nu_m$ 626.46 625.5 | 0.8 1.8 | 390 170 | 20 8.5 | | $r_m = r_K$ 2.291 | $Q'$ 0 | | | $\delta_m$ 5 |
| $\nu_o$ 150.3 | 17 | 4.4 | 6.8 | | $r_o$ 2.507 | $-Q_o'$ 0.3823 | | | $\delta_o$ 16.16 |
| $\nu_{mi}$ 92.9 92.7 92.5 | 2.4 2.3 1.7 | 19 20 27 | 19.2 10.3 10.7 | $x_i$ 0.94012 | $r_m = r_K$ 2.291 | $Q_i'$ 0.1291 | $f(\bar{x_i})$ 2.408 | | $\delta_m = 5$, $\delta_i$ 16.16 |
| $\nu_{oi}$ 29.0 | 4.1 | 3.5 | 20.5 | $x$ 0.85946 | $r_o$ 2.507 | $Q_i'$ 0.1291 | $f(\bar{x})$ 0.9350 | | $\delta_i = \delta_o$ 16.16 |
| $\nu_{mo}$ 25.7 | 3.0 | 4.3 | 5.0 | $x_o$ 0.91420 | $r_o$ 2.507 | $-Q_o'$ 0.3823 | | $\bar{g}(\bar{x_o})$ 2.833 | $\delta_m = 5$, $\delta_o$ 16.16 |
| $\nu_{io}$ 17.9 | 1.9 | 4.7 | 4.0 | $x$ 0.85946 | $r_o$ 2.507 | $-Q_o'$ 0.3823 | | $\bar{g}(\bar{x})$ 2.046 | $\delta_i = \delta_o$ 16.16 |
| $\nu_s$ 0.133 | | large | | | $r_s$ 1 | $-Q_s'$ 0.2641 | | | |
| $\nu_{LT}(m_i)$ (mHz) 1.1 | | | | | $R = r_i$ 2.154 | | | | |

[a] Centroid frequencies $\nu_0$ taken from [22]. [b] Ref. [22]. [c] Quality factors calculated from $Q \equiv \nu_0/(2\Delta)$. [d] Ref. [22]. [e] Definitions of these quantities have been given in section 2.

$10^6$ cm and $Q_s' = -0.2641$ from table 5, a value $U_E = 1.2 \times 10^{52}$ erg can be calculated. This result can be compared the magnetic energy $U_B = \frac{1}{3} R^3 B^2 = 3 \times 10^{47}$ erg ($B_p(\text{tot}) = 1.0 \times 10^{15}$ G has been inserted for $B$). The rotation energy $U_{\text{rot}}$ equals to $U_{\text{rot}} = \frac{1}{2} I \Omega^2 = 3.9 \times 10^{44}$ erg (see table 5). Finally, all these results can be compared with the total emission energy of about $5 \times 10^{46}$ erg, reported by Woods *et al.* [23] for the giant flare of SGR 1806–20 on 27th December 2004.

## 8. OBSERVATIONS ON BLACK HOLE XTE J1550–564

In order to test predictions from section 6 for black hole candidates a set of seven simultaneously observed QPOs is necessary. No such a set is available to my knowledge. Kalemci *et al.* [25], however, reported five simultaneously observed QPO frequencies for the binary system XTE J1550–564 from observations on 16 May, during the decay of the 2000 outburst (set 1 from their tables 1 and 2). Since the two highest frequencies were not detected in that set, we shall use the averaged values for the latter frequencies, found by Miller *et al.* [26] from the outburst in April-May 2000. The used high frequencies can be compared with the corresponding averaged values, obtained by Remillard *et al.* [27] from their analysis of the observations of the 1998–1999 outburst. From the data in their table 1, type B, one may extract: $\nu_i = 277.7$ Hz, $\nu_m = 185.1$ Hz and $\nu_o = 92.6$ Hz. They detected only three low frequency QPOs for the same set type B (see type B spectrum in their figure 1 (bottom left)). The difference between the latter observations and the four low frequency QPOs in set 1, detected by Kalemci *et al.* [25] may again demonstrate the weakness of some QPO signals. As a result, less intense QPOs may easily be missed.

A mass $m_s = 9.61\,m_\odot$ is taken for XTE J1550–564 (see, e.g., [26]). The following quantity can then be calculated

$$\frac{2Gm_s}{c^2} = 2.839 \times 10^6 \text{cm}. \tag{8.1}$$

Data of XTE J1550–564 have been summarized in table 6. In order to obtain the results of table 6, a similar procedure has been applied as before.

Since no values for the parameter $\beta^*$ and the radius $r_s$ of XTE J1550–564 are available, the quantity $\beta^*_{\text{current}}$ cannot be calculated from (6.11). Moreover, the value of the rotation frequency radius $v_s$ is unknown, so that in this case the Lense-Thirring frequency $v_{LT}$ cannot be calculated from (1.8b).

It is noticed that in the derivations of (3.9) and (4.6), leading to $v_{io}$ and $v_{oi}$, respectively, small values for $\delta_i$ and $\delta_o$ have been adopted. Table 6 shows, however, that the found values for $\delta_i$ and $\delta_o$ are not small. In addition, line charges $Q_i$ and $Q_o$ in two tori have been assumed. However, when the charges $Q_i$ and $Q_o$ reside in an inner and outer belt, respectively, the system may also be stable. Thus, the results of table 6 may remain approximately valid.

In addition, it is again noticed that the opposite sign of *all* charges of $Q_s$, $Q_i$ and $Q_o$ given in table 6 is also compatible with the same set of parameters $x$, $r$, $\delta$, $f(\bar{x})$ and $\bar{g}(\bar{x})$ in that table.

Table 6. Data and calculated parameters for XTE J1550–564. See caption of table 2 and text for comments.

| $v_0$ (Hz) | $\Delta$ (Hz) | $Q$ | r.m.s. (%) | $x$ | $R \times 10^6$ (cm) | $Q'$ | $f(\bar{x})$ [c] | $\bar{g}(\bar{x})$ [c] | $\delta$ (°) |
|---|---|---|---|---|---|---|---|---|---|
| $v_s$ ? | | | | | | $Q_s'$ 0.1009 | | | |
| $v_i$ [a] 268 | 56 | 2.4 | 6.2 | | $r_i$ 8.585 | $-Q_i'$ 0.0703 | | | $\delta_i$ 43.97 |
| $v_m$ [a] 188 | 24 | 3.9 | 2.8 | | $r_K$ 9.706 | $Q'$ 0 | | | $\delta_m$ 5 |
| $v_o$ [b] 62.9 | 9.4 | 3.3 | 8.7 | | $r_o$ 11.76 | $Q_o'$ 0.4466 | | | $\delta_o$ 43.97 |
| $v_{mo}$ [b] 8.75 | | | 5.8 | $x_o$ 0.82532 | $r_o$ 11.76 | $Q_o'$ 0.4466 | | $\bar{g}(\bar{x_o})$ 1.7997 | $\delta_m=5$, $\delta_o$ 43.97 |
| $v_{io}$ [b] 5.04 | | | 16.4 | $x$ 0.73000 | $r_o$ 11.76 | $Q_o'$ 0.4466 | | $\bar{g}(\bar{x})$ 1.4348 | $\delta_i=\delta_o$ 43.97 |
| $v_{mi}$ [b] 4.090 | 0.72 | 2.8 | 10.4 | $x_i$ 0.88451 | $r_K$ 9.706 | $-Q_i'$ 0.0703 | $f(\bar{x_i})$ 1.1695 | | $\delta_m=5$, $\delta_i$ 43.97 |
| $v_{oi}$ [b] 0.73 | | | 10.1 | $x$ 0.73000 | $r_o$ 11.76 | $-Q_i'$ 0.0703 | $f(\bar{x})$ 0.4241 | | $\delta_i=\delta_o$ 43.97 |

[a] Centroid frequencies $v_0$, quality factors $Q$ and r.m.s. amplitudes from [26]. [b] Centroid frequencies $v_0$, quality factors $Q$ and r.m.s. amplitudes from [25]. [c] Definitions of these quantities have been given in section 2.

Some additional remarks are in order. For XTE J1550–564 the radius $r_i = 8.58 \times 10^6$ cm is somewhat larger than the corresponding Kepler radius of $7.66 \times 10^6$ cm. Note that the obtained radius for $r_i$ is somewhat larger than $6Gm_s/c^2 = 8.52 \times 10^6$ cm in the Schwarzschild space-time, but much larger than $r_{\text{ISCO}} = 4Gm_s/c^2 = 5.68 \times 10^6$ cm in the extreme limit $|Q_s| = G^{1/2}m_s$ of the Reissner-Nordstrøm space-time.

## 9. OBSERVATIONS ON WHITE DWARFS

Periodic oscillations of different type in white dwarfs have been observed. Warner *et al.* [2] distinguished three types in cataclysmic variable stars: dwarf nova oscillations (DNOs), long period DNOs (lpDNOs) and quasi-periodic oscillations (QPOs). To my knowledge no group of seven simultaneously observed periodic oscillations for any white dwarf has been reported. Therefore, the *gravitomagnetic* precession frequencies (6.6) – (6.9) may not be applicable to these stars. As a possible alternative, the *electromagnetic* precession frequencies from (4.9) and (4.10) will be considered. In that case five precession frequencies will be distinguished: three high frequency DNOs, $v_i$, $v_m$ and $v_o$, and two low frequency periodic oscillations, $v_{io}(em)$ and $v_{oi}(em)$, which may be identified with an lpDNO and a QPO.

As an example, we have selected three simultaneously observed DNOs, one lpDNO and one QPO of the binary dwarf nova VW Hyi, reported by Warner *et al.* [2] and Warner and Woudt [28]. The periods and frequencies of the observed periodic oscillations have been given in table 7, together with their proposed identifications. Periods $P_{lpDNO}$ and $P_{QPO}$ have been chosen, that correspond to the selected $P_{DNO}$s. For completeness sake, the $P_{DNO}$ counterparts have also been given in table 7 (within brackets). The DNOs, lpDNO and QPO have all been observed during the decline of normal and super outbursts. Warner and Woudt [28] selected a time $T = 0$ near the end of the outburst decay where all the light curves of normal and super outbursts coincide. Periodic oscillations were observed at a time $T$ before or after $T = 0$. Their time intervals $<T>$ in days elapsed since $T = 0$ have been added to table 7.

Note that the observed frequencies of VW Hyi are three or four orders of magnitude smaller than their pulsar counterparts in section 7, whereas the r.m.s. amplitudes are about three orders of magnitude smaller. Usually, coherences (i.e., quality factors $Q \equiv v_0/(2\Delta)$ of the DNOs and lpDNOs are more comparable with pulsars, but the coherences of the QPOs are lower.

A mass $m_s = 0.86\,m_\odot$ and a radius $r_s = 6.5 \times 10^8$ cm for VW Hyi have been deduced by Sion *et al.* [29]. The following quantities can then be calculated

$$\frac{2Gm_s}{c^2} = 2.540 \times 10^5 \,\text{cm}, \quad \frac{2Gm_s}{c^2 r_s} = 3.908 \times 10^{-4}. \tag{9.1}$$

The relativistic innermost stable circular orbit $r_{ISCO} = 6Gm_s/c^2 = 7.621 \times 10^5$ cm is much smaller than $r_s = 6.5 \times 10^8$ cm, so that external orbits are not limited by $r_{ISCO}$. With the reported stellar mass $m_s = 0.86\,m_\odot$, the radius $r_m \approx r_K$ can be calculated from (6.5). The result has been given in table 7. A value for rotation period $P_s$ or rotational frequency $v_s$ of VW Hyi has been extracted from observations by Pandel *et al.* [30].

Combination of the electromagnetic precession frequencies $v_{io}(em)$ from (4.9) and $v_{oi}(em)$ from (4.10) yields

$$q = -\frac{v_{io}(em)\,v_i}{v_{oi}(em)\,v_o} = \frac{m_o}{m_i}\,\frac{g(x)}{x\,f(x)} \approx \frac{g(x)^2}{x^3\,f(x)^2}, \tag{9.2}$$

where the approximations $m_o/m_i \approx -Q_o'/Q_i' \approx g(x)/\{x^2 f(x)\}$ have been used (see (6.3) and (6.4)). Note that the last term on the right hand side of (9.2) only depends on $x$. If the angles $\delta_i$ and $\delta_o$ differ from zero value, $g(x)$ and $f(x)$ in (9.2) have to be replaced by $\bar{f}(\bar{x})$ and $\bar{g}(\bar{x})$, respectively. Introduction of the assigned frequencies in the left hand side of (9.2) yields $q = 8.404$. The values for $x$, $\bar{f}(\bar{x})$ and $\bar{g}(\bar{x})$ can then found by iteration as before. All these values have been given in table 7.

In addition, utilizing the assumption $x_i = x_o$ and application of the relation

$$x \equiv r_i/r_o = (r_i/r_m)(r_m/r_o) = x_i x_o = x_i^2 = x_o^2 \quad (9.3)$$

yields the values of $x_i$ and $x_o$. One obtains: $x_i = x_o = x^{1/2} = 0.9218$. Subsequently, the values for $r_i$ and $r_o$ can be calculated. The results have been added to table 7.

In order to calculate the relative charges $Q_i'$, $Q_o'$ and $Q_s'$, additional estimates are necessary. We choose rather arbitrarily

$$\delta_i = \delta_o = 30° \quad \text{and} \quad m_i = 10^{-17} m_\odot. \quad (9.4)$$

Then, from (9.4) and the approximation $m_o/m_i \approx -Q_o'/Q_i' \approx \bar{g}(\bar{x})/\{x^2 \bar{f}(\bar{x})\}$ a value $m_o = 3.1 \times 10^{-17} m_\odot$ follows. Assuming $m_m = 5.9 \times 10^{-17} m_\odot$, one obtains as an estimate for the mass accreting at the boundary layer $m_{accr}$

$$m_{accr} = m_i + m_o + m_m = 10^{-16} m_\odot. \quad (9.5)$$

The boundary layer is thought to be situated between the surface of the white dwarf and the inner edge of the accreting disc (see [30]). Assuming that the life time of the lpDNO and QPO in table 7 is about 0.0073 day ≈ 2.5 $P_{QPO}$ (compare with the time intervals $<T>$ in table 2 of ref. [28]), one finds for the rate of accretion $\dot{m}_{accr}$

$$\dot{m}_{accr} = 5 \times 10^{-12} m_\odot \text{ yr}^{-1}. \quad (9.6)$$

This figure was given by Pandel *et al.* [30] for the rate of accretion at the boundary layer of VW Hyi in the quiescent state. Using (6.3), (6.4), data from (9.4) and needed parameters from table 7, all relative charges $Q_i'$, $Q_o'$ and $Q_s'$ can be calculated. The results have been added to table 7.

Table 7. Data and calculated parameters for dwarf nova VW Hyi. See caption of table 2 and text for comments.

| $<T>$[a] (day) | $P$[a] (s) | $v_0$[d] (mHz) | r.m.s.[a] ×10$^{-3}$ (%) | $x$ | $R \times 10^8$ (cm) | $Q'$ ×10$^{-7}$ | $\bar{f}(\bar{x})$[e] | $\bar{g}(\bar{x})$[e] | $\delta$ (°) |
|---|---|---|---|---|---|---|---|---|---|
| 0.37 | $P_{DNO}$ 13.69 | $v_i$ 73.05 | 4.1 | | $r_i$ 9.952 | $-Q_i'$ 1.45 | | | $\delta_i$ 30 |
| 0.37 | $P_{DNO}$ 20.86 | $v_m$ 47.94 | 11.7 | | $r_m \approx r_K$ 10.80 | $Q'$ 0 | | | |
| 0.37 | $P_{DNO}$ 38.28 | $v_o$ 26.12 | 4.3 | | $r_o$ 11.71 | $Q_o'$ 4.55 | | | $\delta_o$ 30 |
| | $P_s$[b] 62 | $v_s$[b] 16 | 6,200[b] | | $r_s$ 6.5 | $Q_s'$ 2.85 | | | |
| −0.11 | ($P_{DNO}$ 20.31) $P_{lpDNO}$[c] 84.44 | $v_{io}$(em) 11.84 | 1.4 1.7[c] | $x$ 0.8497 | $r_o$ 11.71 | $-Q_i'$ 1.45 $Q_o'$ 4.55 | | $\bar{g}(\bar{x})$ 1.965 | $\delta_i, \delta_o$ 30 |
| 0.30 | ($P_{DNO}$ 20.94) $P_{QPO}$ 254 | $v_{oi}$(em) 3.94 | 3.6 9.4 | $x$ 0.8497 | $r_o$ 11.71 | $-Q_i'$ 1.45 $Q_o'$ 4.55 | $\bar{f}(\bar{x})$ 0.8653 | | $\delta_i, \delta_o$ 30 |

[a] Time intervals $<T>$, periods $P$ and r.m.s. amplitudes have been taken from [28]. [b] Data from [30]. [c] Data from [2]. See for $<T>$ ref. [28]. [d] Centroid frequencies $v_0$, calculated from the periods in column on the left. [e] Definitions of these quantities have been given in section 2.

An estimate of the electric field $E$ at the surface of the white dwarf can now be calculated from $Q_s'$

$$E = \frac{Q_s}{r_s^2} = \frac{Q_s' G^{1/2} m_s}{r_s^2} = 3.0 \times 10^5 \text{ statvolt.cm}^{-1} = 8.9 \times 10^9 \text{ V.m}^{-1}, \quad (9.7)$$

where $m_s = 0.86\, m_\odot$ and $r_s = 6.5 \times 10^8$ cm have been inserted. Contributions to $E$ from the charges $Q_i$ and $Q_o$ have been neglected (compare with the more complete expressions (2.28) and (2.29)).

There is an indication that an electric field of the order of magnitude given by (9.7) has already been observed. The very broad absorption Lyman-alpha feature measured by, e.g., Sion *et al.* [31] may be attributed to the so-called *linear* Stark effect, obtained by first-order perturbation theory (see, e.g., Landau and Lifshitz. [32]). In order to illustrate the relevant parameters of this effect, we consider the four states of the second energy level ($n = 2$) of the hydrogen atom: an $l = 0$ state (2s) and three $l = 1$ states (2p) with $m = -1, 0, 1$, respectively. All these states possess the same energy $E_2^0$. In the presence of an external electric field $E$ the latter energy level is split up in three different energy levels: $E_2^0 - 3ea_0E$, $E_2^0$ and $E_2^0 + 3ea_0E$, where $e$ is the absolute charge of an electron and $a_0$ is the Bohr radius. This splitting is known as the *linear* Stark effect, since the energy shifts are linear in $E$. (The first energy level of the hydrogen atom with $n = 1$, $E_1^0$, is not affected by $E$). Thus, in the presence of an electric field the energy of the original Lyα line is shifted upwards and downwards, respectively, by an amount of $3ea_0E$. The energy of the original Lyα line is $E_2^0 - E_1^0 = hc/\lambda \approx \tfrac{3}{8} e^2/a_0$, whereas the wavelength equals to $\lambda \approx (8hca_0)/(3e^2) = 1215.0$ Å (or more precisely $\lambda = 1215.7$ Å)). From the splitting of the Lyα line the magnitude of the electric field can then be calculated to be

$$E = \frac{hc\,\Delta\lambda}{6ea_0\lambda^2} = 6.7 \times 10^4 \text{ statvolt.cm}^{-1} = 2.0 \times 10^9 \text{ V.cm}^{-1}, \quad (9.8)$$

where the value $\Delta\lambda = 76$ Å has been extracted from [31] (see their figure 2). It is noticed that for the calculated field $E$ of (9.8) the *linear* Stark effect is greater than the so-called *quadratic* Stark effect, that can be calculated by second-order perturbation theory. Summing up, the result (9.8) can be regarded to be in reasonable agreement with the value of (9.7). In view of the assumptions involved, however, both results of (9.7) and (9.8) are uncertain.

For VW Hyi an absolute value for $B_p(gm) = 9.1 \times 10^8$ G can be calculated from (1.4), using the values $m_s = 0.86\, m_\odot$, $r_s = 6.5 \times 10^8$ cm, $\nu_s = 0.016$ Hz and $\beta = +1$. Since it is thought that the magnetic field of VW Hyi is much weaker, we will assume that $\beta^* = 0$ (compare with (5.2)), for convenience sake. Introducing the other necessary data from table 7 into (6.11), the following value for $\beta^*_{\text{current}}$ can then be calculated

$$\beta^*_{\text{current}} = -1 + \{2.85(Q_s) - 5.49(Q_i) + 5.97(Q_o)\} \times 10^{-7} = -1 + 3.33 \times 10^{-7}. \quad (9.9)$$

In view of the uncertain value of $\beta^*$, the value of $\beta^*_{\text{current}}$ can only be regarded as an estimate. The value of $\beta^*_{\text{current}}$ resembles the values found for the rapidly rotating pulsars in this work and the asymptotic value $\beta^*_{\text{current}} = -1$, suggested for millisecond pulsars in ref. [8].

It is again noticed that the opposite sign of *all* charges of $Q_s$, $Q_i$ and $Q_o$ in table 7 is also compatible with the same set of parameters $x$, $r$, $\delta$, $\bar{f}(\bar{x})$ and $\bar{g}(\bar{x})$ in that table. Otherwise stated, the sign of any of the charges $Q_s$, $Q_i$ and $Q_o$ is not yet known.

Finally, an attempt has been made to extract a set of seven frequencies for white dwarf VW Hyi from the data reported by Warner *et al.* [2] and Warner and Woudt [28]. Using the high frequency DNOs from table 7 and identifying four other frequencies with one of the gravitomagnetic precession frequencies (6.6)–(6.9), it appeared possible to find such a set parameters, analogous to those of the discussed pulsars. The calculated value of $Q_s'$, however, is about 250, a value that seems unacceptable high. Therefore, the electromagnetic precession frequencies (4.9) and (4.10) seem to be the better choice. However, in view of many uncertainties (e.g., identification of frequencies, rate of accretion, magnitude of the electric field) the results of table 7 are only an estimate.

## 10. DISCUSSION AND CONCLUSIONS

All high frequency quasi-periodic oscillations (QPOs) in this work are attributed to three different circular tori: an inner torus with charge $Q_i$, a torus with electrically neutral mass $m_m$ and an outer torus with charge $Q_o$. The corresponding QPO frequencies $\nu_i$, $\nu_m$ and $\nu_o$ of the binary pulsars SAX J1808.4–3658, XTE J1807–294 and IGR J00291+5934, the soft gamma repeater SGR 1806–20, the black hole XTE J1550–564 and the white dwarf VW Hyi are the three highest frequencies, following the sequence: $\nu_i > \nu_m > \nu_o$. The values of the corresponding radii follow the sequence: $r_i < r_m < r_o$. The frequencies $\nu_i$, $\nu_m$ and $\nu_o$, and the radii $r_i$, $r_m$ and $r_o$ for all these stars are presented in sections 7–9 and are summarized in table 8. Note that the considered stars strongly differ with respect to the values of the rotation frequencies $\nu_s$, frequencies $\nu_i$, $\nu_m$ and $\nu_o$, and radii. Therefore, it is striking that it appears possible to describe the observed high frequency QPOs by the same set of frequencies $\nu_i$, $\nu_m$ and $\nu_o$.

Contrary to the relativistic precession model proposed by Stella and Vietri [3], and Stella, Vietri and Morsink [4], in this work the so-called *lower* kHz frequency, $\nu_l$, is identified with the orbital frequency $\nu_m$ of (2.27) (in all discussed examples the frequency $\nu_m$ nearly coincides with the Kepler frequency $\nu_K$). In the former model the so-called *upper* kHz frequency, $\nu_u$, is attributed to the frequency $\nu_m$ in our notation. Whereas in their model a nearly circular orbit for the mass current is assumed, we only consider circular orbits in our model.

Table 8. Summary of frequencies $\nu_s$, $\nu_i$, $\nu_m$ and $\nu_o$ and radii $r_s$, $r_i$, $r_m$ and $r_o$ from tables 2–7 for a number of strongly different stars.

| Star | $\nu_s$ (Hz) | $\nu_i$ (Hz) | $\nu_m$ (Hz) | $\nu_o$ (Hz) | $r_s$ ×10⁶ (cm) | $r_i$[d] ×10⁶ (cm) | $r_m$ ($r_K$) ×10⁶ (cm) | $r_o$[d] ×10⁶ (cm) |
|---|---|---|---|---|---|---|---|---|
| Pulsars[a] | | | | | | | | |
| SAX J1808.4–3658 | 401 | 682.4 | 503.3 | 189 | 1 | 2.22 (2.16) | 2.64 (2.65) | 3.10 (5.09) |
| XTE J1807–294 | 191 | 544.9 | 358 | 134 | 1 | 2.69 (2.51) | 3.32 (3.32) | 3.47 (6.40) |
| IGR J00291+5934 | 598.88 | 66.3 | 5.3 | 0.71 | 1 | 43.8 (10.2) | 55.1 (55.1) | 59.0 (211) |
| SGR 1806–20 | 0.133 | 1837 | 625.5 | 150.3 | 1 | 2.15 (1.12) | 2.29 (2.29) | 2.51 (5.93) |
| Black hole XTE J1550–564[b] | ? | 268 | 188 | 62.9 | | 8.58 (7.66) | 9.71 (9.71) | 11.8 (20.1) |
| White dwarf VW Hyi[c] | 0.016 | 0.07305 | 0.04794 | 0.02612 | 650 | 995 (815) | 1080 (1080) | 1171 (1618) |

[a] Adopted mass, $m_s = 1.4\, m_\odot$, yielding $r_{ISCO} = 6\,Gm_s/c^2 = 1.24 \times 10^6$ cm in the Schwarzschild space-time. [b] $m_s = 9.61\, m_\odot$, yielding $r_{ISCO} = 6\,Gm_s/c^2 = 8.52 \times 10^6$ cm. [c] $m_s = 0.86\, m_\odot$, yielding $r_{ISCO} = 6\,Gm_s/c^2 = 0.76 \times 10^6$ cm. [d] Within brackets: Kepler radii corresponding to the values $r_i$ and $r_o$ deduced from our model.

At present, it is difficult to decide which high frequency QPO may be attributed to the orbital frequency $v_m$ of (2.27), closely related to the Kepler frequency. For the SGR 1806–20 the highest QPO frequency 1837 Hz leads to a Kepler radius of $1.12 \times 10^6$ cm (for SGR 1806–20 the angular momentum contribution to the radius can be neglected), smaller than the relativistic innermost circular orbit $r_{ISCO} = 6Gm_s/c^2 = 1.24 \times 10^6$ cm in the Schwarzschild space-time. In any case, identification of the 1837 Hz QPO with $v_i$ in our model leads to an acceptable value for the inner radius $r_i = 2.14 \times 10^6$ cm (see section 7.4 and table 8).

It can be seen from table 8 that all calculated radii in table 8 are larger than the radius $r_{ISCO} = 6Gm_s/c^2$. In some cases it is found that the radius $r_i$ of the inner torus does not differ very much from the corresponding Kepler radius (see table 8). Note that for IGR J00291+5934 the radius $r_m$ is much larger than the stellar radius $r_s$. Future studies of the time dependence of $v_m$ and $r_m$ may give us more information about the evolution of, e.g., the torus with mass $m_m$.

The calculated radii $r_o$ of the outer torus in table 8 are systematically smaller than the corresponding Kepler radii. These findings indicate that the charge dependent contribution to $v_o$ of (6.2) cannot be neglected in general. As a consequence, the validity of (6.4) will be limited. The results calculated from the gravitomagnetic precession frequencies (6.6) and (6.7), however, are not affected by this complication, since the frequency $v_o$ has been substituted as an empirical parameter in these equations.

In the considered examples, the high frequency QPOs $v_i$ and $v_m$ usually possess large quality factors $Q$ ($Q \equiv v_0/(2\Delta)$, see section 7) and r.m.s. amplitudes, whereas the quality factors of the QPO frequency $v_o$ are often small. As a result, $v_o$ may escape detection. However, in the studied pulsars the r.m.s. amplitudes of the frequency $v_o$ are not so marginal. An possible explanation of this behaviour may be as follows. When the radii of $r_s$ and $r_m$ do not differ much, there is less space between the surface of the star and the torus with neutral mass $m_m$. The quality factor $Q$ and r.m.s. amplitude of $v_i$ may then be relatively high. Likewise, when the radii of $r_i$ and $r_o$ do not differ much, the quality factor $Q$ and r.m.s. amplitude of $v_m$ may also be high. Outside radius $r_o$ more space may be available, resulting into a low quality factor $Q$ but a rather high r.m.s. amplitude. Although usually two prominent high frequency QPOs are observed in Z sources and atoll sources (see, e.g., van der Klis [1, § 2.9]), three high frequency QPOs were reported by Jonker *et al.* [33] for the atoll sources 4U 1608–52, 4U 1728–34 and 4U 1636–53.

In table 9 the assigned low frequency QPOs $v_{mo}$, $v_{io}$, $v_{mi}$ and $v_{oi}$ from (6.6)–(6.9) for four pulsars and one black hole from tables 2–7 have been summarized. In addition, the relative charges $Q_i'$, $Q_o'$ and $Q_s'$ ($Q_i'$ is defined by $Q_i' \equiv (G^{1/2}m_s)^{-1}Q_i$ and so on), and the parameters $\beta^*$ and $\beta^*_{current}$ have been added.

For all pulsars in table 8 and 9 seven frequencies have simultaneously been observed, whereas the precession model given in refs. [3, 4] only accounts for three basic QPO frequencies. Apart from the Lense-Thirring precession (see, e.g., (1.9)), in our model seven QPO frequencies are predicted (see section 6).

Numerous attempts have been undertaken to find correlations between the QPO frequencies (see, e.g., [1 (§ 2.7), 4, 15, 19–21]). In order to compare the correlation between the QPO frequencies $v_{mi}$ and $v_i$ in our model, equation (6.8) can be rewritten as follows

$$\frac{v_{mi}}{v_i} = Q_i' \frac{2Gm_s}{c^2 r_m} x_i f(x_i) \cos\delta_m \cos\delta_i, \quad x_i \equiv r_i/r_m. \tag{10.1}$$

This relation shows that the predicted ratio $v_{mi}/v_i$ depends on several largely independent parameters: the relative charge $Q_i'$, the radii $r_i$ and $r_m$ ($f(x_i)$ is fixed by $x_i \equiv r_i/r_m$) and the angles $\delta_m$ and $\delta_i$. Thus, only when the time behaviour of a sufficient number of parameters is known, the behaviour of another parameter can be tested.

Table 9. Summary of frequencies $\nu_{mo}$, $\nu_{io}$, $\nu_{mi}$ and $\nu_{oi}$, relative charges $Q_s'$, $Q_i'$ and $Q_o'$ from tables 2–6 for pulsars and a black hole, and parameters $\beta^*$ and $\beta^*_{current}$.

| Star | $\nu_{mo}$ (Hz) | $\nu_{io}$ (Hz) | $\nu_{mi}$ (Hz) | $\nu_{oi}$ (Hz) | $Q_i'$ | $Q_o'$ | $Q_s'$ | $\beta^*$ [c] | $\beta^*_{current}$ [c] |
|---|---|---|---|---|---|---|---|---|---|
| Pulsar[a] | | | | | | | | | |
| SAX J1808.4 −3658 | 73.6 | 46.27 | 15.02 | 4.97 | −0.24 | +1.62 | +0.33 | ≈ 0 | −0.50 |
| XTE J1807 −294 | 38.3 | 7.65 | 3.63 | 1.67 | −0.17 | +0.82 | +0.26 | ≈ 0 | −0.74 |
| IGR J00291 +5934 | 0.0223 | 0.012 | 0.052 | 0.0430 | −0.30 | +1.77 | +0.44 | ≈ 0 | −0.56 |
| SGR 1806–20 | 25.7 | 17.9 | 92.7 | 29.0 | +0.13 | −0.38 | −0.26 | +125 | +1275 |
| Bl. hole XTE J1550 −564[b] | 8.75 | 5.04 | 4.090 | 0.73 | −0.070 | +0.45 | +0.10 | | |

[a] Used mass: $m_s = 1.4$. [b] $m_s = 9.61\, m_\odot$. [c] See equation (6.11) and section 7.

Further, according to the relativistic precession model given in refs. [3, 4], a relation will exist between a third low frequency QPO, the Lense-Thirring frequency $\overline{\nu_{LT}^-}$ of (1.9), and the highest QPO frequency, the *upper* frequency $\nu_u$. When $\nu_u$ is approximated by the Kepler frequency $\nu_u = (2\pi)^{-1}(Gm_s/r_u^3)^{1/2}$, the following relation between $\overline{\nu_{LT}^-}$ from (1.9) and $\nu_u$ for a nearly circular orbit is obtained

$$\overline{\nu_{LT}} = \frac{4\nu_s r_s^2}{5c^2}\frac{Gm_s}{R^3} = \frac{16\pi^2 \nu_s r_s^2}{5c^2}\nu_u^2, \tag{10.2}$$

where $R = r_u$ is the Kepler radius. In order to test (10.2), several authors (see, e.g., refs. [1, (§ 2.7–§ 2.8 (review)), 4, 15]) tried to find correlations between low frequency QPOs and the highest frequency, the upper frequency $\nu_u$. It was found that low frequency QPOs often displayed an approximate dependence on the squared upper frequency $\nu_u$, but in ref. [1, § 2.8] it was stated: "the gravitomagnetic effect has not yet been detected with certainty in any system".

Comparison with our model shows related dependencies of *all* low frequency QPOs, i.e, $\nu_{mo}$, $\nu_{io}$, $\nu_{mi}$ and $\nu_{oi}$ on the high frequency QPOs $\nu_i$ or $\nu_o$. For example, the low frequency QPO $\nu_{mi}$ (see (10.1)) depends on the product of the quantities $\nu_i = (2\pi)^{-1}(Gm_s/r_i^3)^{1/2}$ (see relation (2.15)) and $Gm_s/r_m$. Results for the Lense-Thirring frequency $\nu_{LT}$ according to (1.8b) for circular tori with the smallest radii, $r_i$ and $r_m$, have been given in tables 2–5. At present, no definite identifications can be made, however. Owing to a factor two in the expression for $\overline{\nu_{LT}^-}$ of (10.2) compared with the frequency $\nu_{LT}$ from (1.8b), combined with the influence of deviations from a circular orbit, the value for the Lense-Thirring frequency in the model of refs. [3, 4] will be larger than $\nu_{LT}$ from (1.8b).

It appears that all relative charges $Q_s' \equiv (G^{1/2}m_s)^{-1}Q_s$ for the pulsars in table 9 possess a rather high absolute value of about 0.32. As already pointed out in section 7.1, the value of $Q_s'$ may be somewhat lower when the value of $m_s$ is higher. In that case, the product of $m_s$ and $Q_s'$, and the gravitomagnetic precession frequencies of (6.6)–(6.9) remain the same. High values for $Q_s'$ have been considered by several authors, e.g., by Ghezzi [10].

He concluded that charged stars in hydrostatic equilibrium have a somewhat larger mass and radius than the uncharged ones. The latter effect is due to Coulomb repulsion. Therefore, the pulsar radius $r_s$ may be larger than the standard value used in table 8. An alternative choice for $r_s$ does not change the predicted values of the QPO frequencies $v_i$, $v_m$, $v_o$, $v_{mo}$, $v_{io}$, $v_{mi}$ and $v_{oi}$ of section 6 ($r_s$ does not occur in the formulas of these frequencies), but it affects the predicted value of the parameter $\beta^*$ of (6.11).

No detailed mechanism is given in this work, how a compact star remains stable with such large amounts of charge. This subject requires further investigation. It has been shown in section 2, however, that an equilibrium may be formed between the charge $Q_s$ in the star and the charges $Q_i$ and $Q_o$ in the tori. Then, the electric fields from $Q_s$, $Q_i$ and $Q_o$ at the radii $r_i$ and $r_o$ compensate each other (see comment to equations (2.8) and (2.20)). The net electric field at the centre of the star is also zero. Although the radial electric field at the equator of the star $E_{eq}(tot)$ is far from being zero (see (2.29)), it has been shown in section 2 that equilibrium between the charges may exist. Accretion may disturb this equilibrium, so that the charge of the star $Q_s$ will change (see comment to (2.25)). In addition, nuclear burning at the surface of a pulsar may also affect the magnitude of the charge $Q_s$ (see discussion of SAX J1808.4–3658 in section 7.1). Furthermore, a large electric field $\mathbf{E}_p(tot)$ in the direction of $\mathbf{s} = \mathbf{\Omega}_s/\Omega_s$ is predicted at the poles (see (2.28)). Along this direction no charge equilibrium needs to exist. As a result, polar outflow of charge (*jets*) will occur. This subject will receive more attention in the future.

Although the described Coulomb interaction between the charges $Q_s$, $Q_i$ and $Q_o$ will usually dominate, the magnetic field in stars itself will also lead to a monopole charge $Q_{mon}$ in the star, as has been pointed out by Michel and Li [34, § 4]. In addition, they deduced a model with a dome of charge over each magnetic pole and a torus of charge with the opposite sign in the equatorial zone ("*dome and torus*" model). This model has previously been compared with our gravitomagnetic approach in ref. [8]. Extensions of the "*dome and torus*" model have been given (see [8] and refs. [32] and [33] therein). In first order the magnitude of the monopolar contribution to $Q_s'$ equals to $Q_{mon}' = (G^{1/2}m_s)^{-1}Q_{mon} = -2/15\,(2\pi v_s r_s/c)^2$ for $\beta = +1$ (see [8]). But even the pulsar with the largest rotation frequency $v_s = 598.88$ Hz in our sample, IGR J00291+5934, only yields a modest contribution to $Q_s'$, i.e., $Q_{mon}' = -2.1 \times 10^{-3}$. For that reason, $Q_{mon}'$ has been neglected in this work.

A detailed discussion of the value of the parameter $\beta^*$ for binary, isolated, short- and long-period pulsars has previously been given by Biemond [8]. Following this paper, a value of $\beta^* \approx 0$ is assumed for the *rapidly* rotating binary pulsars SAX J1808.4–3658, XTE J1807–294 and IGR J00291+5934 in this work, whereas values for $\beta^*_{current}$ ranging from – 0.50 up to – 0.74 are found (see table 9). For comparison, only for the isolated, millisecond pulsar B1821–24 ($v_s = 328$ Hz) an *observational* value $\beta^* = 2\times 10^{-5}$ could be extracted from electron cyclotron resonance spectral features, whereas a value of $\beta^*_{current} \approx -1$ for this pulsar has been proposed (see discussion in ref. [8]). For the *slowly* rotating SGR 1806–20 the dominating magnetic field from the torus with charge $Q_i'$ may cause a large magnetic field at the poles of the star, resulting into the large value $\beta^* = +125$. Moreover, this strong field may have induced the large value of $\beta^*_{current} = +1275$ in the star in the past (see discussion in section 7.4).

In the deduction of the QPO frequencies $v_i$, $v_m$, $v_o$, $v_{mo}$, $v_{io}$, $v_{mi}$ and $v_{oi}$ (see section 6) and the Lense-Thirring frequency $v_{LT}$ (see (1.8b)) it has been assumed that the angles $\delta_i$, $\delta_m$ and $\delta_o$ are small (see, however, the results of, e.g., XTE J1807–294 discussed in section 7.2). In fact, throughout this work the normal unit vectors $\mathbf{n}_i$, $\mathbf{n}_m$ and $\mathbf{n}_o$ (perpendicular to the directions of the planes of the three tori, see text and figure 3) are assumed to be approximately parallel to the direction of the rotation axis $\mathbf{s}$ of the star. Only for the Lense-Thirring frequency $v_{LT}$ of (1.9) another limiting case has been considered: $\mathbf{s}$ perpendicular to, e.g., $\mathbf{n}_m$. For the discussed pulsars results for the Lense-Thirring frequency $v_{LT}$ according to (1.8b) have been given in tables 2–5.

The results for the black hole XTE J1550–564 and the considered pulsars show a close relationship. For example, the values of $2Gm_s/(c^2 r_i)$ for XTE J1550–564 and SAX J1808.4–3658 are 0.33 and 0.19, respectively. These values are characteristic for the general relativistic regime. Moreover, the QPO frequencies of both type of stars can be described by the same set of deduced frequencies: $v_i$, $v_m$, $v_o$, $v_{mo}$, $v_{io}$, $v_{mi}$ and $v_{oi}$ (see section 6).

For white dwarfs the value of $2Gm_s/(c^2 r_i)$ is much smaller than for pulsars and black holes. For example, for the dwarf nova VW Hyi $2Gm_s/(c^2 r_i)$ is $2.6 \times 10^{-4}$ (see (9.1) and table 7). The high frequency QPOs of the white dwarfs have been explained by the same model applied to the pulsars and the black hole. However, the low frequency QPOs might be attributed to the purely electromagnetic precession frequencies $v_{io}$(em) (see (4.9)) and $v_{oi}$(em) (see (4.10)). In section 9 predictions are compared with data from the dwarf nova VW Hyi.

## 11. SUMMARY

A new model for the explanation of quasi-periodic oscillations (QPOs) in pulsars, black holes and white dwarfs is presented. In our approach a new key ingredient, *charge*, is introduced. In the presented model it is adopted that accreting pulsars, black holes and white dwarfs all bear an electric charge $Q_s$, whereas charges $Q_i$ and $Q_o$ are present in an inner and outer torus around the star, respectively. The signs of the charges and $Q_s$ and $Q_o$ are assumed to be the same, whereas the signs of $Q_s$ and $Q_i$ are thought to be opposite. In addition, a third torus with an electrically neutral mass $m_m$, is assumed to be present between the other tori. It is assumed that all orbits of $Q_i$, $m_m$ and $Q_o$, are circular, with radii $r_i$, $r_m$ and $r_o$, respectively. The frequencies of the currents, due to the charges $Q_i$ and $Q_o$, are calculated from classical mechanics and Coulomb's law, and are denoted by $v_i$ (see (6.1)) and $v_o$ (see (6.2)), respectively. The frequency $v_m$ of the mass current from $m_m$ is closely related to the standard Kepler frequency (see (6.5)). Since the considered stars strongly differ with respect to the values of the rotation frequency $v_s$, frequencies $v_i$, $v_m$ and $v_o$, and other properties, it is striking that it appears possible to describe the observed high frequency QPOs by the same set of frequencies $v_i$, $v_m$ and $v_o$.

Another key ingredient in this work is *a special interpretation of the gravitomagnetic theory*, deduced from general relativity. This version of the theory [5–8] predicts the existence of a dipolar magnetic field for all rotating stars, embodied in the so-called Wilson-Blackett law. The deviation of the observed magnetic field from the predicted field can be described by the parameter $\beta^*$ (see (5.2)). For the proposed model with charges $Q_i$ and $Q_o$ in two tori an expression for $\beta^*$ (see (5.7)) has been deduced, connecting the rotational frequency $v_s$ of the star to the QPO frequencies $v_i$ and $v_o$. Applications of the relation $\beta^*$ have been given.

For the explanation of the low frequency QPOs in pulsars and black holes the special interpretation of the gravitomagnetic theory is also essential. The latter approach results in the deduction of four new *gravitomagnetic* precession frequencies: $v_{mo}$, $v_{io}$, $v_{mi}$ and $v_{oi}$ (see equations (6.6–6.9)), which have been identified with observed low frequency QPOs. Thus, apart from the Lense-Thirring frequency $v_{LT}$ (see equations (1.8b) and (1.9)), seven QPO frequencies are predicted in our model in this case: $v_i$, $v_m$, $v_o$, $v_{mo}$, $v_{io}$, $v_{mi}$ and $v_{oi}$.

Predictions of the proposed model are compared with observed high frequency and low frequency QPOs of the pulsars SAX J1808.4–3658, XTE J1807–294, IGR J00291 +5934 and SGR 1806–20. The results seem to be compatible with the presented model. Moreover, charge flow near the pole of pulsars may explain the observed frequency drift of burst oscillations. Likewise, charge flow at the equator of SAX J1808.4–3658 may be the cause of the enigmatic 410 kHz QPO. Furthermore, results comparable to the four pulsars are obtained for the black hole XTE J1550-564.

The high frequency QPOs of a white dwarf may be explained by the same model, as has been applied to the pulsars and the black hole. The low frequency QPOs might be attributed to two purely *electromagnetic* precession frequencies: $v_{io}$(em) of (4.9) and $v_{oi}$(em) of (4.10). Predictions are compared with data from the dwarf nova VW Hyi.

Summing up, in the presented three tori model the high frequency QPOs of widely different stars, like four pulsars, one black hole and one white dwarf are all described by the same set of three orbital frequencies $v_i$, $v_m$ and $v_o$. The low frequency QPOs for pulsars and black hole may be attributed to four *gravitomagnetic* precession frequencies: $v_{mo}$, $v_{io}$, $v_{mi}$ and $v_{oi}$, whereas the low frequency QPOs of a dwarf nova may be attributed to two *electromagnetic* precession frequencies: $v_{io}$(em) and $v_{oi}$(em). Although the calculated results, like radii $r_i$, $r_m$ and $r_o$, and relative charges $Q_s'$, $Q_o'$ and $-Q_i'$ ($Q'$ is defined by $Q' \equiv (G^{1/2} m_s)^{-1} Q$) can only be regarded as a first order approximation, these parameters may be essential in our understanding of the quasi-periodic oscillations. In conclusion, the new model seems to be in agreement with more observations than previously proposed alternatives.

## ACKNOWLEDGEMENT

I would like to thank Albert van der Beek for solving computer problems and his wife Irene Biemond for programming help. The technical help of my son Pieter in publishing this paper is also gratefully acknowledged. I thank many others for interest. Finally, I thank my wife Nel for her continuous support since the start of this work in 1979.